\documentclass[11pt,a4paper]{article}
\pdfoutput=1
\usepackage{sint}
\usepackage{amsmath,amsfonts}
\usepackage{graphicx,cite,url}
\usepackage{amssymb}
\usepackage{mathrsfs}  
\usepackage{color}
\usepackage{macros_static}
\usepackage{booktabs,dcolumn}
\usepackage{hyperref}
\makeatletter
\def\@xcmidrule{\ifx\@tempa\cmidrule\vskip-\@thisrulewidth
     \global\@lastruleclass=\@ne\else
     \ifx\@tempa\morecmidrules\vskip \cmidrulesep
     \global\@lastruleclass=\@ne\else
     \vskip \belowrulesep\global\@lastruleclass=\z@\fi\fi
     \ifnum0=`{\fi}}
\makeatother

\begin{document}

\begin{titlepage}

\begin{flushright}
\small{
DESY 12-026 \\
Edinburgh 2011/41\\
HU-EP-12/05 \\
LPT-Orsay/12-28 \\
MKPH-T-12-08 \\
MS-TP-12-01 \\
SFB/CPP-12-10 \\
}
\end{flushright}

\begin{center}
{\Large\bf
Parameters of Heavy Quark Effective Theory \vspace{0.5cm}\\
from $\nf=2$ lattice QCD} 
\vskip 1.25cm
\vbox{\includegraphics[width=2.8cm]{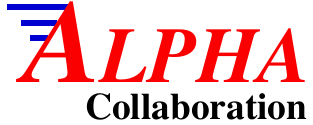}}
\vskip 1.25cm
{
Beno\^it~Blossier$^a$, Michele~Della~Morte$^b$, Patrick~Fritzsch$^c$, Nicolas~Garron$^d$, 
Jochen~Heitger$^e$, Hubert~Simma$^f$, Rainer~Sommer$^f$, Nazario~Tantalo$^{g,h}$
}
\vskip 0.5cm
\it {\footnotesize
$^{\scriptstyle a}$
LPT, 
CNRS et Universit\'e Paris-Sud XI,
B\^atiment 210, 91405~Orsay~Cedex, France\\[0.2em]
$^{\scriptstyle b}$ Universit\"at~Mainz, Institut~f\"ur~Kernphysik, Becherweg~45, 55099~Mainz, Germany \\[0.2em]
$^{\scriptstyle c}$ Institut f\"ur Physik, Humboldt Universit\"at, Newtonstr. 15, 12489~Berlin, Germany\\[0.2em]
$^{\scriptstyle d}$ Tait institute, School of Physics and Astronomy, 
University of Edinburgh, Edinburgh, EH9 3JZ, UK\\[0.2em]
$^{\scriptstyle e}$ Universit\"at M\"unster, Institut f\"ur Theoretische Physik, 
Wilhelm-Klemm-Str. 9, 
48149 M\"unster, Germany \\[0.2em]
$^{\scriptstyle f}$ NIC, DESY, Platanenallee 6, 15738 Zeuthen, Germany \\[0.2em]
$^{\scriptstyle g}$ Dip. di Fisica, Universit\`a di Roma 'Tor Vergata', Via della Ricerca Scientifica 1, 
I-00133 Rome, Italy\\[0.em]
$^{\scriptstyle h}$ INFN, Sez. di Roma 'Tor Vergata', Via della Ricerca Scientifica 1, 
I-00133 Rome, Italy
}
\vskip 1cm
{\bf Abstract}
\vskip 0.1ex
\end{center}
We report on a non-perturbative determination of the parameters of the lattice Heavy Quark Effective 
Theory (HQET) Lagrangian and of the time component of the heavy-light axial-vector current
with $\nf=2$ flavors of massless dynamical quarks.
The effective theory is considered at the $\minv$ order, 
and the heavy mass $\mh$ covers a range 
from slightly above the charm to beyond the beauty region.
These HQET parameters are needed to compute, for example, 
the b-quark mass, the heavy-light spectrum and decay constants
in the static approximation and to order $\minv$ in HQET.
The determination of the parameters is done non-perturbatively.
The computation reported in this paper uses 
the plaquette gauge action and two different 
static actions for the heavy quark described by HQET. For the light-quark action
we choose non-perturbatively $\Or(a)$-improved Wilson fermions.


\vskip 2.0ex

\eject
\vfill
\eject

\end{titlepage}

\section{Introduction}
\label{s:intro}

Particle physics enters very exciting times with the start of collecting data
at the Large Hadron Collider.  Two ways are explored to probe New Physics (NP):
either performing a direct search of new particles (e.g. at ATLAS and CMS)
from the electroweak scale up to the \TeV~scale, or studying rare decays (e.g. at
LHCb).
The latter give rise to a very rich set of constraints on NP scenarios
because they are either mediated by quantum loops, in which 
high energy particles circulate (that is the case, for instance, in flavor changing
neutral currents), or by new current structures (in decays occuring at tree level).
B-mesons offer a highly interesting and rich set of (rare) decay channels, such as 
$b \to s \gamma$, which arises in the standard model only from penguin diagrams, 
and hence puts strong bounds on NP scenarios. The analysis of inclusive decays 
such as $B \to X_s \gamma$, in particular in the framework of a heavy quark expansion, 
strongly depends on the knowledge of the b-quark mass, $\mb$. In tests of the CKM 
mechanism, some tensions exist at the moment between $\sin 2\beta$, obtained from 
the golden mode $B \to J/\Psi K_s$, and the CKM matrix element $V_{ub}$ extracted 
from the $B \to \tau \nu$ leptonic decay~\cite{sinbeta}. 
The theoretical input of 
the latter is the B-meson decay constant $\fB$, whose uncertainty is $\sim$ 10\%. 
Currently, there is also a 3-$\sigma$ discrepancy between 
$V^{B \to \tau \nu}_{ub}$ and $V^{B \to \pi l \nu}_{ub}$. 
Although it would be surprising if this difference were due 
to a significant underestimate of $f_B$, 
reducing its error is an important task of lattice QCD.

Lattice QCD will enable us to compute $\mb$ and $\fB$ 
with a precision comparable to the one of forthcoming experimental measurements from high luminosity collisions. 
Nevertheless, a delicate issue for those extractions is how to get a satisfying control on the
cut-off effects. Indeed, the Compton length of the b-quark is smaller than the
typical finest lattice spacing of simulations in large volumes 
(used to compute hadronic quantities).
Several strategies have been explored in the literature to circumvent this
two-scale difficulty~\cite{Thacker:1990bm, Lepage:1992tx, ElKhadra:1996mp,
Guagnelli:2002jd, Aoki:2001ra, Christ:2006us, Blossier:2009hg},
see~\cite{Davies:2012qf} for a recent review.
The ALPHA collaboration has proposed to use the framework of Heavy Quark Effective Theory
(HQET)~\cite{Eichten:1987xu, Eichten:1989zv} with a non-perturbative determination 
of the couplings~\cite{Heitger:2003nj}.  
Its implementation at the first order in $\minv$ has successfully been applied 
in the quenched approximation~\cite{DellaMorte:2006cb,Blossier:2010jk,Blossier:2010vz,
Blossier:2010mk}.  In this paper we will report on our effort to realize our
HQET program for $\nf=2$ flavors of dynamical quarks, leaving the
phenomenological results for forthcoming papers. 
Here, in particular, we present
our determination of the couplings of the effective theory regularized on the
lattice.

We use the same notations as in~\cite{Blossier:2010jk}. The HQET Lagrangian
density at the leading (static) order is given by 
\bes
\lag{stat}(x) = \heavyb(x) \,D_0\, \heavy(x)\;.
\ees
A bare quark mass $\mhbarestat$ has to be added to the energy levels $\Estat$
computed with this Lagrangian to obtain the physical ones.  For example, the mass of the
B-meson in the
static approximation is given by
\bes 
\mB = \Estat + \mhbarestat \;.
\ees
At the classical level $\mhbarestat$ is simply the (static approximation of
the) b-quark mass, $\mbeauty$, but in the quantized lattice formulation it has
to further compensate a divergence, an inverse power of the lattice spacing.
Including the $\minv$ terms, the HQET Lagrangian reads%
\footnote{%
The precise definitions of $D_0$, ${\bf D}^2$ and $
{\boldsymbol\sigma}\!\cdot\!{\bf B}$ can be found in~\cite{Blossier:2010jk}.}
\bes
\lag{HQET}(x) &=&  \lag{stat}(x) - \omegakin\Okin(x)
        - \omegaspin\Ospin(x)  \,, \\[2.0ex]
  \Okin(x) &=& \heavyb(x){\bf D}^2\heavy(x) \,,\quad
  \Ospin(x) = \heavyb(x){\boldsymbol\sigma}\!\cdot\!{\bf B}\heavy(x)\,.
\ees
At this order, two other unknown parameters appear in the Lagrangian,
$\omegakin$ and $\omegaspin$. Our normalization is such that the {\em
classical} values of the coefficients are $\omegakin=\omegaspin=1/(2\mh)$.
In order to compute the decay constant of a heavy-light meson, one needs the
time component of the axial-vector heavy-light current $A_0$.  At the lowest
order of the effective theory, the current is form-identical to the
relativistic one. 
At the $\minv$ order, it is enough to add only one term to the static current 
(because we are only interested in zero-momentum correlation functions, see~\cite{Sommer:2010ic})
\bes
 \label{eq:ahqet}
 \Ahqet(x)&=& \zahqet\,[\Astat(x)+  \cah{1}\Ah{1}(x)]\,.
\ees
In this work we present our non-perturbative determination of the parameters
$\mhbare$ (the generalization of $\mhbarestat$ to the $\minv$ order),
$\lnzahqet$, $\cahqet$, $\omegakin$, and $\omegaspin$ at values of the
lattice spacing relevant for the computation of hadronic observables.  
These parameters allow us to compute, for example,
the spectrum of heavy-light mesons and heavy-light decay constants.  As we
explain in detail in the remainder  of this paper, the basic idea is to match,
in a small volume, a few observables expanded in the effective theory at finite
lattice spacing to their non-perturbative continuum values determined in QCD
at a given renormalization group invariant (RGI) heavy quark mass $M$.
Collecting the observables in a vector $\Phi$, the matching equation reads
\be
\label{eq:match_L}
\Phi^{\rm HQET}(L,M,a) = \Phi^{\rm QCD} (L,M,0) \;,
\ee
where $L$ is the space extent of the lattice and $a\to0$ taken in QCD. Expanding the
left hand side of \eq{eq:match_L} at a given order of the 
inverse heavy quark mass defines a set of HQET parameters. 
We remind the reader that such an order-by-order treatment~\cite{Heitger:2003nj}
is part of the very definition of HQET.

The remainder of the text is organized as follows: in~\sect{s:s2} we summarize
the strategy of the computation, in~\sect{s:s3} we give the details of its
implementation, the final results can be found in~\sect{s:s4} and~\sect{s:s5}
contains our conclusions.  Some definitions are relegated
to~\ref{a:observables}, in~\ref{a:tuning} we explain how we tuned the
parameters of the simulations and how we performed the renormalization of the
QCD quantities, whereas~\ref{a:sim_details} contains more technical
information about the dynamical fermion runs.

\section{Strategy}
\label{s:s2}

The computation reported here is done along the lines
of~\cite{Blossier:2010jk}.  For completeness, we repeat here the basic
ingredients but refer the reader to this work for more detailed explanations.
We start by the simulations of QCD in a small volume with space extent 
$L_1\approx 0.4 \fm$ at four different values of the lattice spacing.  
We consider $\nf=2$ dynamical light quarks that we tune to be massless 
and a quenched heavy (valence) quark that we simulate at nine different mass values,
such that the lightest mass is around the charm mass and the heaviest mass is above the b-quark mass.
We compute five renormalized observables that we extrapolate to the continuum
\bes
\Phi_i^{\rm QCD}(L_1,M,0) = \lim_{a\to0} \Phi_i^{\rm QCD}(L_1,M,a)\;,\qquad i=1,\ldots, 5.
\ees
The definitions of these observables can be found in~\cite{Blossier:2010jk} and
in~\ref{a:observables}. Here we just mention that $\Phi_1$ and $\Phi_2$ are 
finite volume versions of the heavy-light meson mass and the logarithm of the
decay constant, respectively (up to kinematic factors).  $\Phi_3$ is sensitive
to the correction of the heavy-light current. Finally, $\Phi_4$ and $\Phi_5$
are proportional to the kinetic and magnetic corrections, respectively.  At the
next-to-leading order of the effective theory (i.e. keeping the static and the
$\minv$ terms) these observables can be expressed in terms of the five parameters 
discussed in~\sect{s:intro}, which we cast into a vector 
\be 
\omega=\left(\mhbare,\,\lnzahqet,\,\cahqet,\,\omegakin,\,\omegaspin\right)^{\rm t}\;.
\ee
More precisely, we write the $\minv$ expansion of the observables in the following way%
\footnote{For example, in the case of the static-light meson mass, $\eta$
is proportional to the static energy, and this equation is simply
the finite volume version of $\mB=\Estat+\mhbare$\,.}%
:  
\bes
\label{eq:phihqet}
 \Phi^{\rm HQET} = \phistat + \phimat\,\omega\,\;.
\ees
The entries of the five-component vector $\phistat$  and the five-by-five
block-diagonal matrix $\phimat$ are computed by performing a series of 
numerical simulations of HQET 
at fixed $L=L_1$ and for various lattice spacings $a$. 
A more explicit form of~\eq{eq:phihqet} can be found in~\ref{a:observables}.
As anticipated in the introduction, the matching condition that we impose
is~\eq{eq:match_L} with $L=L_1$:
\be
\label{eq:match_L1}
\Phi^{\rm HQET}(L_1,M,a) = \Phi^{\rm QCD}(L_1,M,0) \;.
\ee
Solving this equation defines the HQET parameters that we call
$\tilde\omega(M,a)$
\bes
\label{eq:omega_tilde}
\tilde\omega(M,a) = 
\phimat^{-1}(L_1,a) 
\Big(
\Phi^{\rm QCD}(L_1,M,0) -  \phistat(L_1,a)
\Big)\,.
\ees
They are the bare couplings of the theory and, as such, can be determined
by finite volume matching conditions. By imposing eq.~\eqref{eq:match_L1},
the parameters~$\tilde\omega_i$ become functions of $M$, but this heavy quark
mass dependence comes entirely from $\Phi^{\rm QCD}$.
We then perform another set of simulations of the
effective theory in a larger volume of space extent $L_2=2L_1$.  The
observables in this volume are then simply obtained by taking the continuum
limit of eq.~\eqref{eq:phihqet} in which we insert the parameters
$\tilde\omega(M,a)$ computed in the previous step:
\bea
\label{eq:Phi_L2}
 \Phi^{\rm HQET}(L_2,M,0)
&=& \lim_{a\to0}
\Big[
 \phistat(L_2,a) + \phimat(L_2,a)\, \tilde\omega(M,a)
\Big]\,. 
\eea
Our formulation of the theory and the non-perturbative determination
of $\tilde\omega$ guarantees that all divergences, including those of
order $1/a\,,\; 1/a^2$ are cancelled, and that the limit $a\to0$ exists.
Next, the parameters at larger lattice spacings (to be used in
large volume) are obtained by inverting~\eq{eq:phihqet} with $L=L_2$, 
\bes
\omega(M,a) = 
\phimat^{-1}(L_2,a) 
\Big(
\Phi^{\rm HQET}(L_2,M,0) -  \phistat(L_2,a)
\Big)\;.
\ees
Finally, in the last step we perform an interpolation (or, in one case, a slight extrapolation)
in the inverse bare coupling $\beta=6/g_0^2$ and obtain $\omega(M,a)$ at 
exactly those values 
of the lattice spacing used in our large volume 
simulations\cite{Blossier:2011dk}. \\

\section{Numerical application}
\label{s:s3}
%
\subsection{Continuum extrapolation of the QCD observables}
\label{s:s3.1}
For the QCD simulation, we use the plaquette gauge action and non-perturbatively 
$\Or(a)$-improved clover fermions~\cite{Jansen:1998mx} for $\nf=2$ flavors of 
dynamical quarks with Schr\"odinger functional (SF) boundary conditions. 
We have four different
values of the lattice spacing ($\lesssim 0.02\fm$),  $\beta$ varying in the
range $6.16-6.64$, and the number of lattice points per space direction being
$L_1/a=20,24,32,40$.  The physical volume is kept fixed by imposing for the
Schr\"odinger functional coupling the value $\gbsq(L_1/2)=2.989$, 
which corresponds to $L_1\approx 0.4 \fm$.  
For each resolution $L_1/a$ we tune the dimensionless RGI heavy quark mass such that 
\bes
z=L_1M \in\{4\,, 6\,, 7\,, 9\,, 11\,, 13\,, 15\,, 18\,, 21\} \;.
\ees
With this choice, $M$ varies approximately from $2\GeV$ to $10\GeV$.  
More details about their renormalization and
about the tuning of the bare coupling and quark masses can be found in~\ref{a:tuning}. 
For the run parameters we refer the reader to~\ref{a:sim_details}.  Finally, we
also implement tree-level improvement, following exactly the procedure
described in Appendix D of~\cite{Blossier:2010jk}.

\begin{figure}[tb]
\begin{center}
\begin{tabular}{cc}
\includegraphics[width=0.45\textwidth]{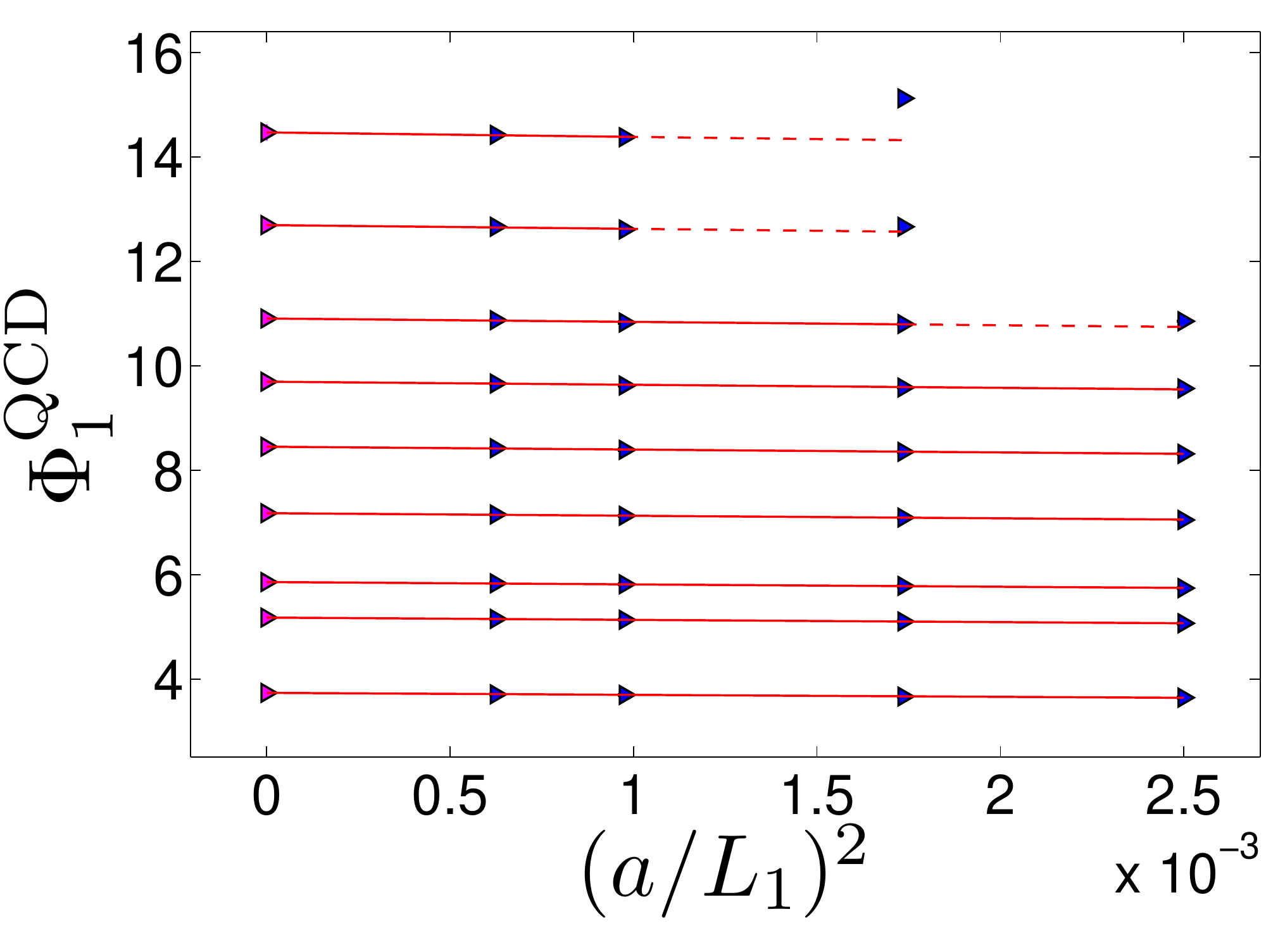}&
\qquad
\includegraphics[width=0.45\textwidth]{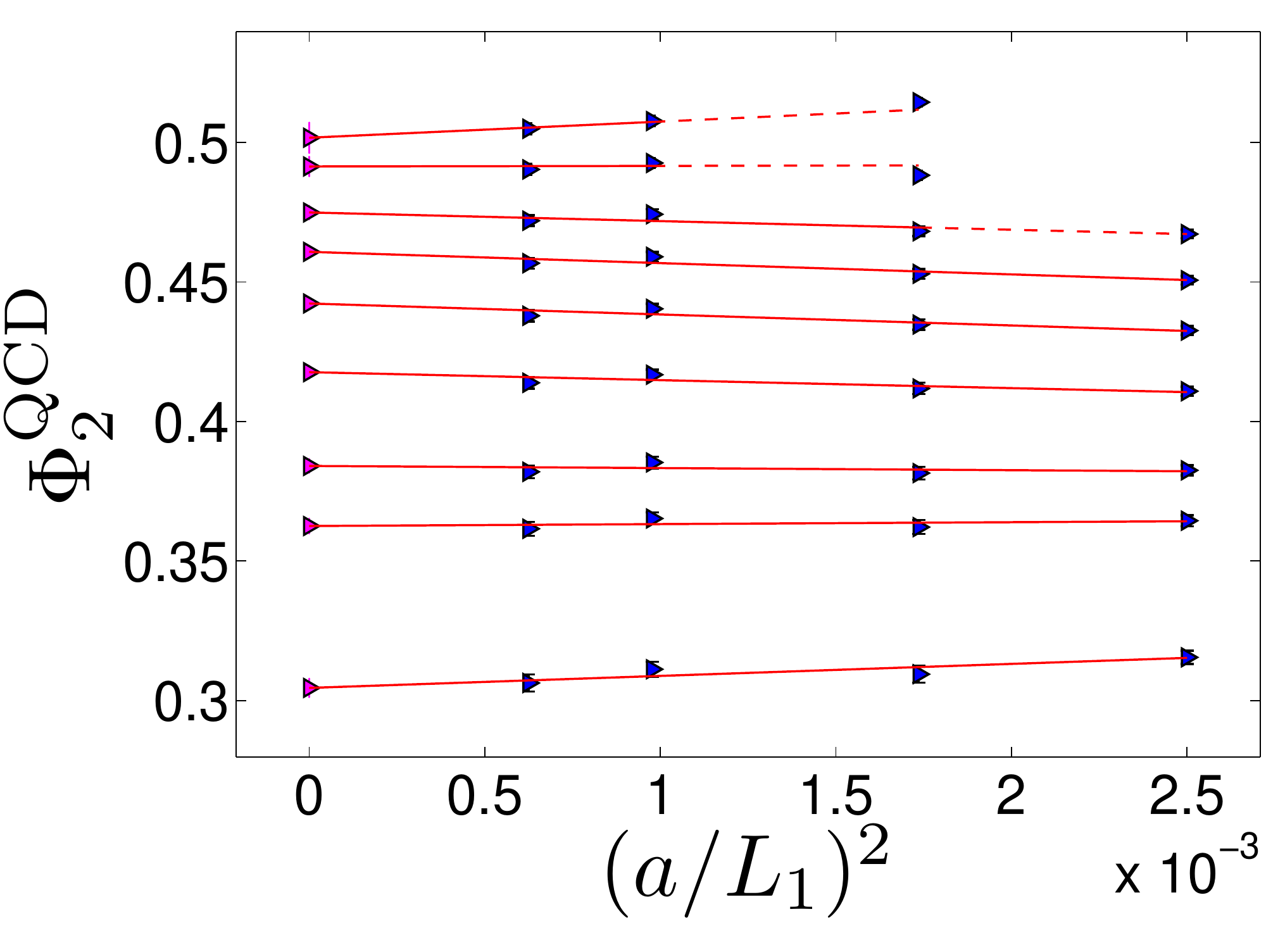}\\[0.5em]
\includegraphics[width=0.45\textwidth]{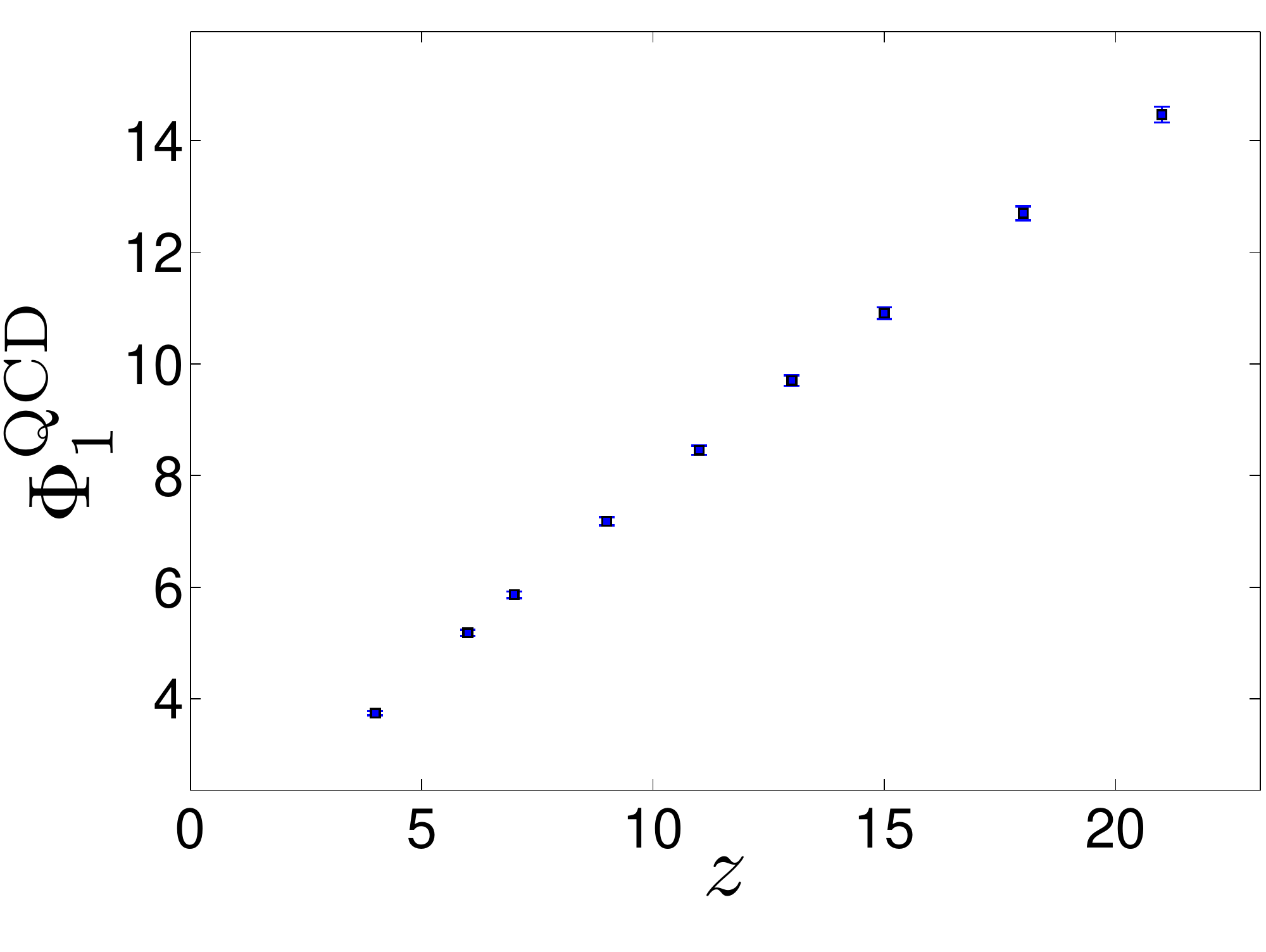}&
\qquad
\includegraphics[width=0.45\textwidth]{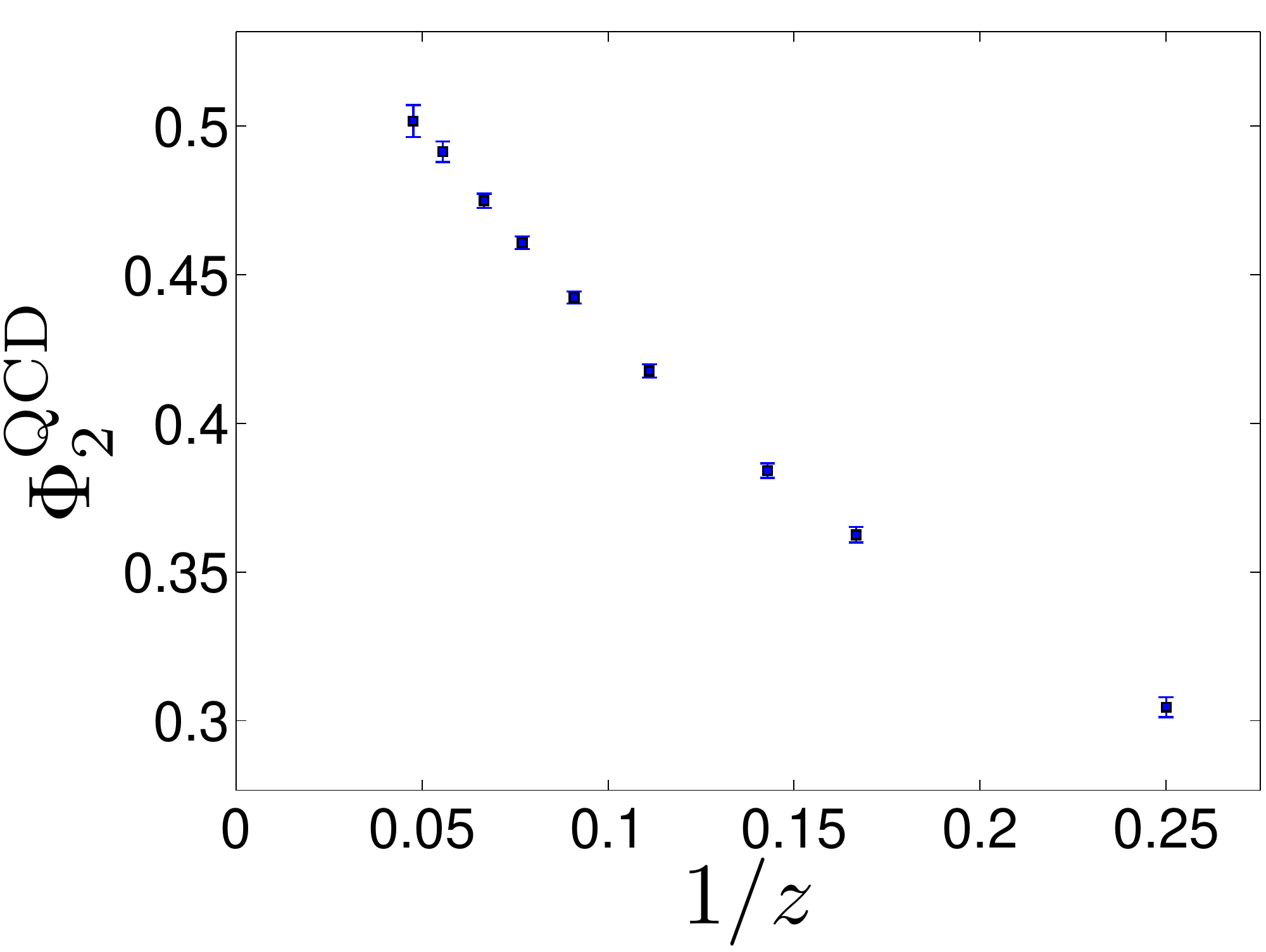}\\
\end{tabular}
\caption{\footnotesize
On the top panels we show the continuum extrapolation of two 
QCD observables where we used a global fit to parametrize the mass dependence
of the slopes in $(a/L)^2$. 
The observable shown on the left is proportional to a finite-volume heavy-light meson mass 
and the one on the right to the logarithm of a finite-volume decay constant.
On the lower panels we show the heavy quark mass dependence 
of these observables ($z= L_1M$).
}
\label{fig:PhiQCD}
\end{center}
\end{figure}

Our strategy for the continuum extrapolation differs somewhat from our previous
work.  The discretization effects can be important for our heaviest masses.
In particular, for the simulations where $L_1/a \le 24$, we might have
noticeable contributions of order $(a/L)^n$ with $n>2$.  We take
advantage of the fact that we have various (and quite different) heavy quark
masses: since the cut-off effects are smooth functions of $a/L_1$ and $z$,
we perform a global fit of the form 
\bes
\label{eq:globalfit}
\Phi^{\rm QCD}(L,M,a) = \Phi^{\rm QCD}(L,M,0) 
\Big[1 + (a/L_1)^2 \; (A +B \, z +C\,z^2) \Big] \;,
\ees
using only the data points such that $aM < 0.7$, as motivated in
\cite{Heitger:2003ue,Kurth:2001yr}.  Note that the two last terms 
in \eq{eq:globalfit} are proportional to $(a /L_1 ) \times (aM)$ 
and ${(a M)^2}$.  
For each observable, $\Phi^{\rm QCD}_i$, the parameters of the fit are the 
nine different values $\Phi_i^{\rm QCD}(L,M,0)$ and $A_i,\,B_i, \,C_i$.
As an illustration we show
the results of the fit of $\Phi_1$ and $\Phi_2$ in \fig{fig:PhiQCD}.
We have checked that different fit ans\"atze (e.g. adding cubic lattice artefacts) 
give consistent results 
for both the central values and the errors, and that results are
also compatible with the standard approach where the slope in $(a/L)^2$
is not constrained.

\subsection{Subtraction of the static part}
\label{s:s3.2}
The effective theory is first simulated in the small volume of space extent $L_1$,
which is tuned to be the same as in QCD\footnote{
In practice this tuning is done with a certain precision, which translates into a small 
error on the various observables. Since the static quantities are very precise 
this error is dominating for some of them and was taken into account
as explained in~\ref{a:tuning}.
}. We use five different lattice
spacings, such that $L_1/a = 6, 8, 10, 12, 16$ (but for the continuum
extrapolation we discard the coarsest point).  The corresponding $\beta$-values lie in
the range $5.26 - 5.96$.  For the static quark we use two different lattice
actions, HYP1 and HYP2, 
which are known to lead to small statistical errors~\cite{DellaMorte:2005yc}. Once
again the light quarks are tuned to be massless, both in the valence and in the
sea sectors.  More details on the simulations
can be found in the appendices.  This set of simulations serves to compute the quantities
$\phistat(L_1,a)$ and $\phimat(L_1,a)$ in \eq{eq:omega_tilde}. 

\begin{figure}[tb]
\begin{center}
\begin{tabular}{cc}
\includegraphics[width=0.45\textwidth]{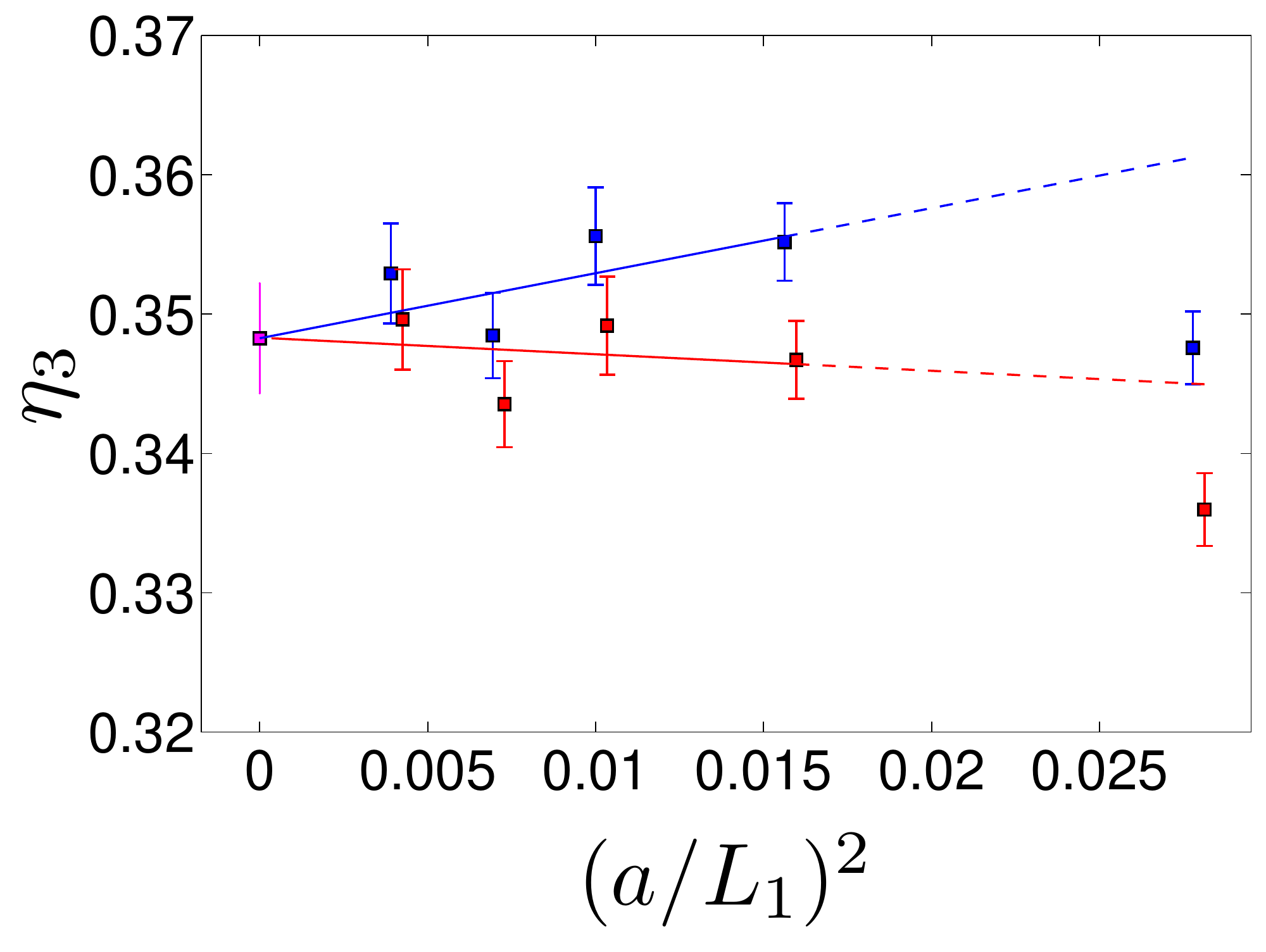}&
\quad
\includegraphics[width=0.45\textwidth]{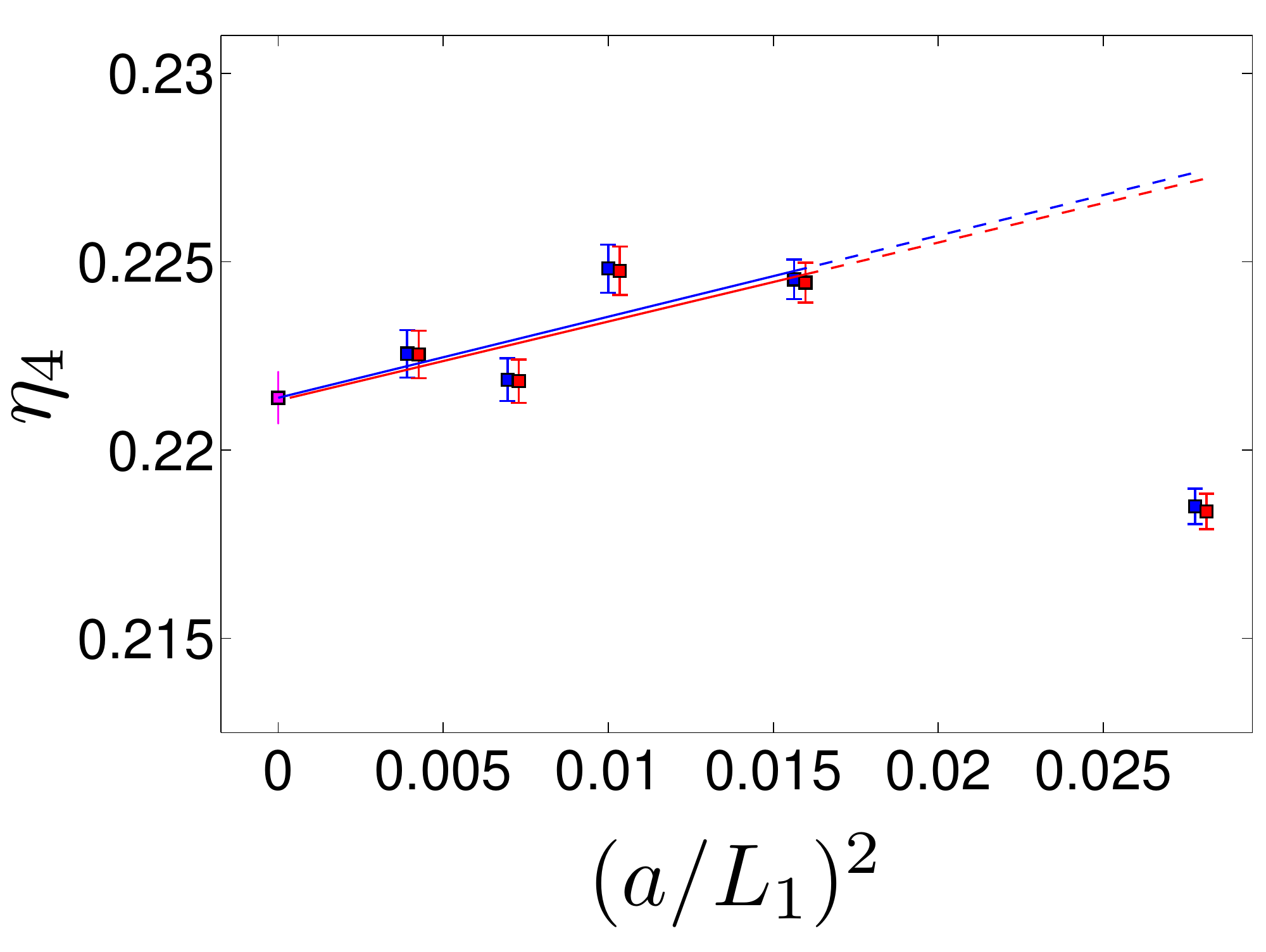}
\end{tabular}
\end{center}
\vspace{-2em}
\caption{\footnotesize
Continuum extrapolation of 
$\eta_3(L_1,a)$ and $\eta_4(L_1,a)$, 
which are the static part of $\Phi_3(L_1,M,a)$ and $\Phi_4(L_1,M,a)$ respectively.
The two different colors represent the static action: blue for HYP2 
and red for HYP1 (slightly shifted to the right).
}
\label{fig:PhiRA1stat}
\end{figure}

For $i=3,4,5$, the static contributions of $\Phi_i$ are $\eta_i$.
They have a well defined continuum limit and $\eta_5 = 0$. Thus 
we compute
\bes
\eta_i(L_1,0) = \lim_{a\to0} \eta_i(L_1,a)\,, \qquad i=3,4 \;,
\ees
and show the continuum extrapolations in~\fig{fig:PhiRA1stat}.  They are done
linearly in $(a/L_1)^2$, which is justified by two reasons:
(i) The quantity $\eta_3$ contains the time-component of the static axial 
      current; we improve this current using the 1-loop value of $a\castat$ 
      computed in~\cite{Grimbach:2008uy}. This is sufficient since the 
      improvement term has almost no numerical influence: even setting
      $\castat=0$ gives compatible results. 
(ii) The quantity $\eta_4$ is constructed from SF boundary-to-boundary correlators, 
      which are already $\Or(a)$-improved in our setup, cf.~\ref{a:sim_details}.

\begin{figure}[tb]
\begin{center}
\begin{tabular}{cc}
\includegraphics[width=0.45\textwidth]{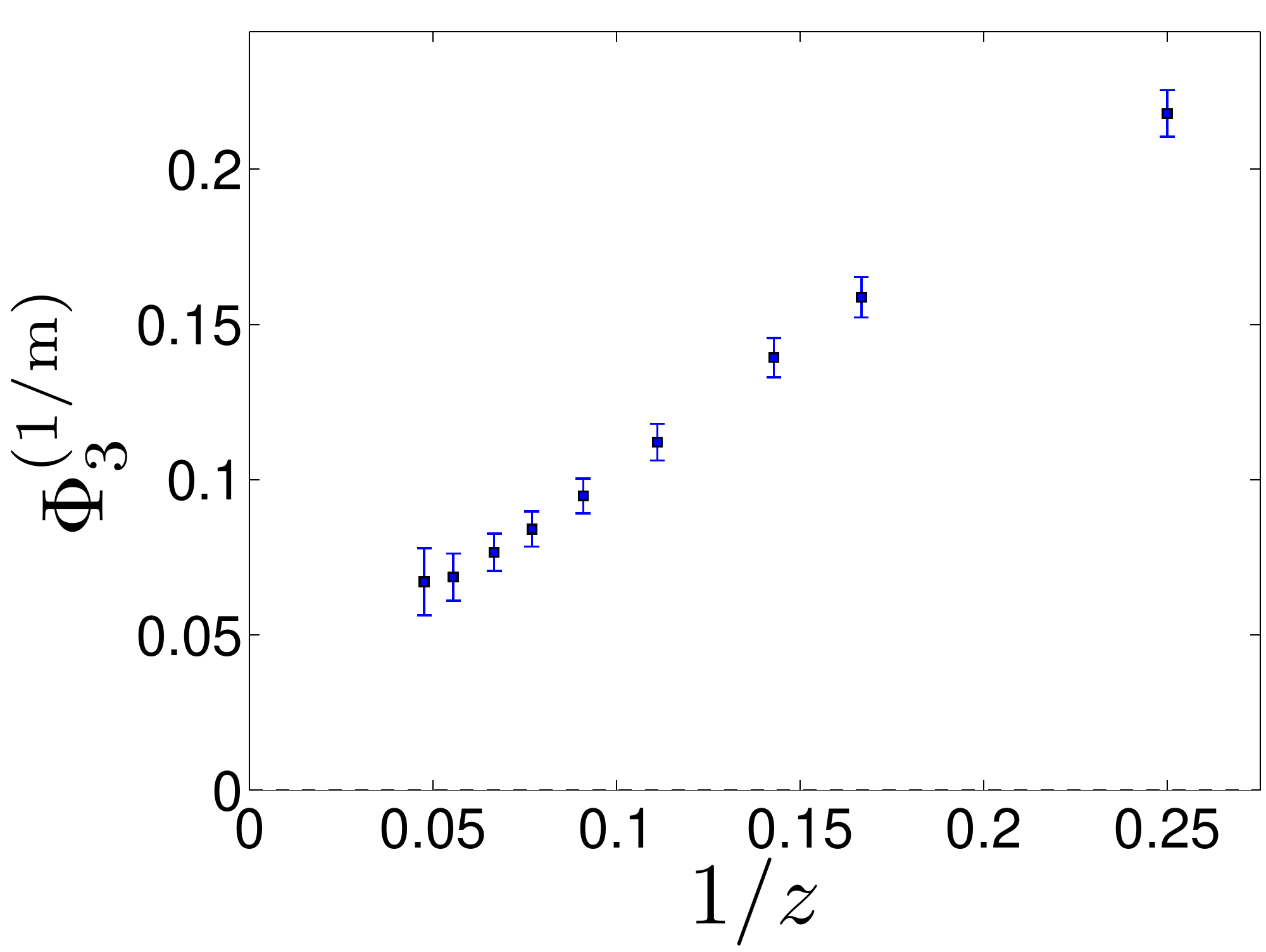}&
\quad
\includegraphics[width=0.45\textwidth]{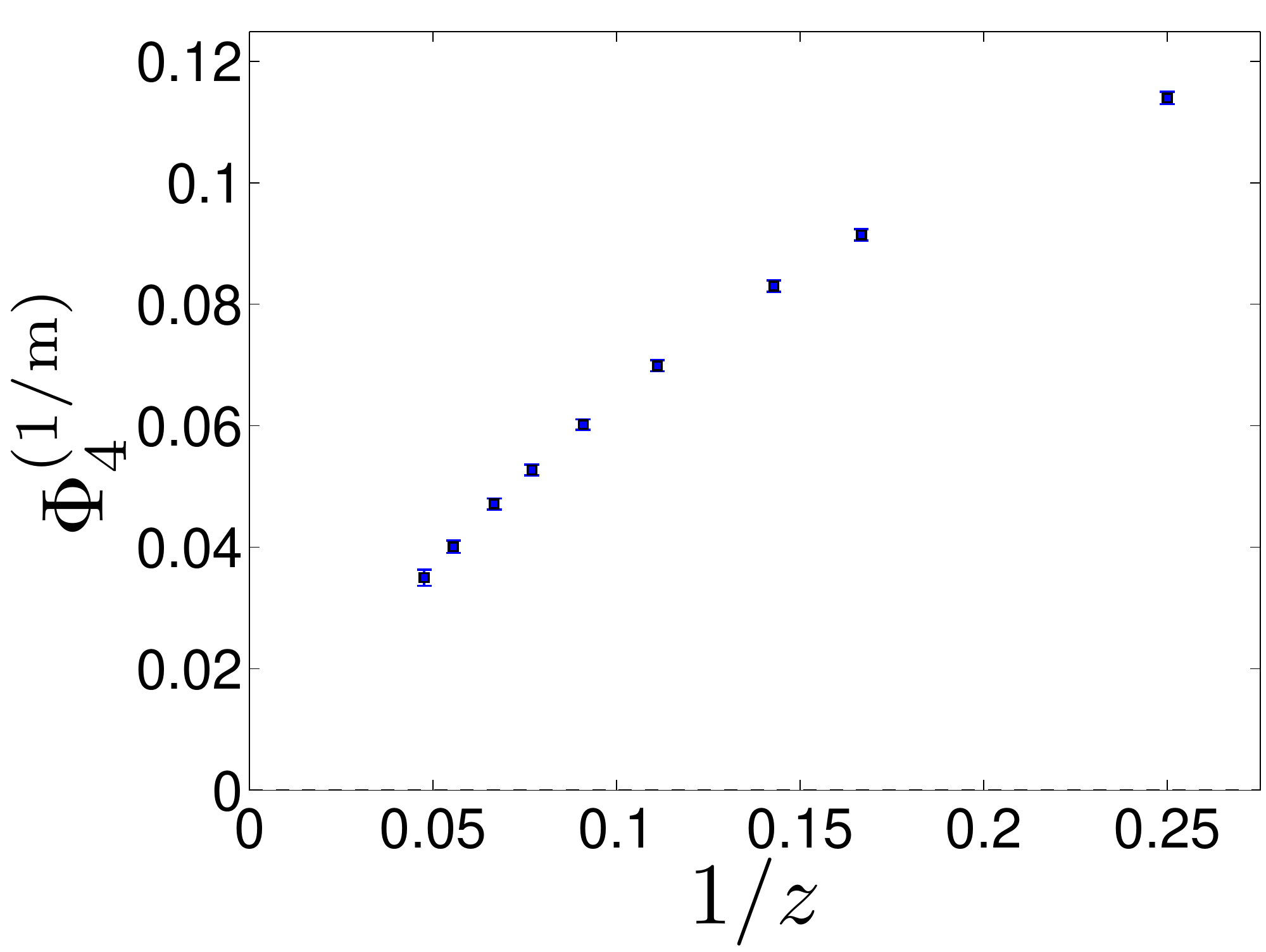}
\end{tabular}
\end{center}
\vspace{-2em}
\caption{\footnotesize
Mass behavior of $\Phi_{3,4}^{\first}(L_1,M,0)$. As explained in the text, 
these quantities are expected to vanish in the static limit $1/z=0$.
}
\label{fig:Phi341om}
\end{figure}

\begin{figure}[tb]
\begin{center}
\includegraphics[width=0.45\textwidth]{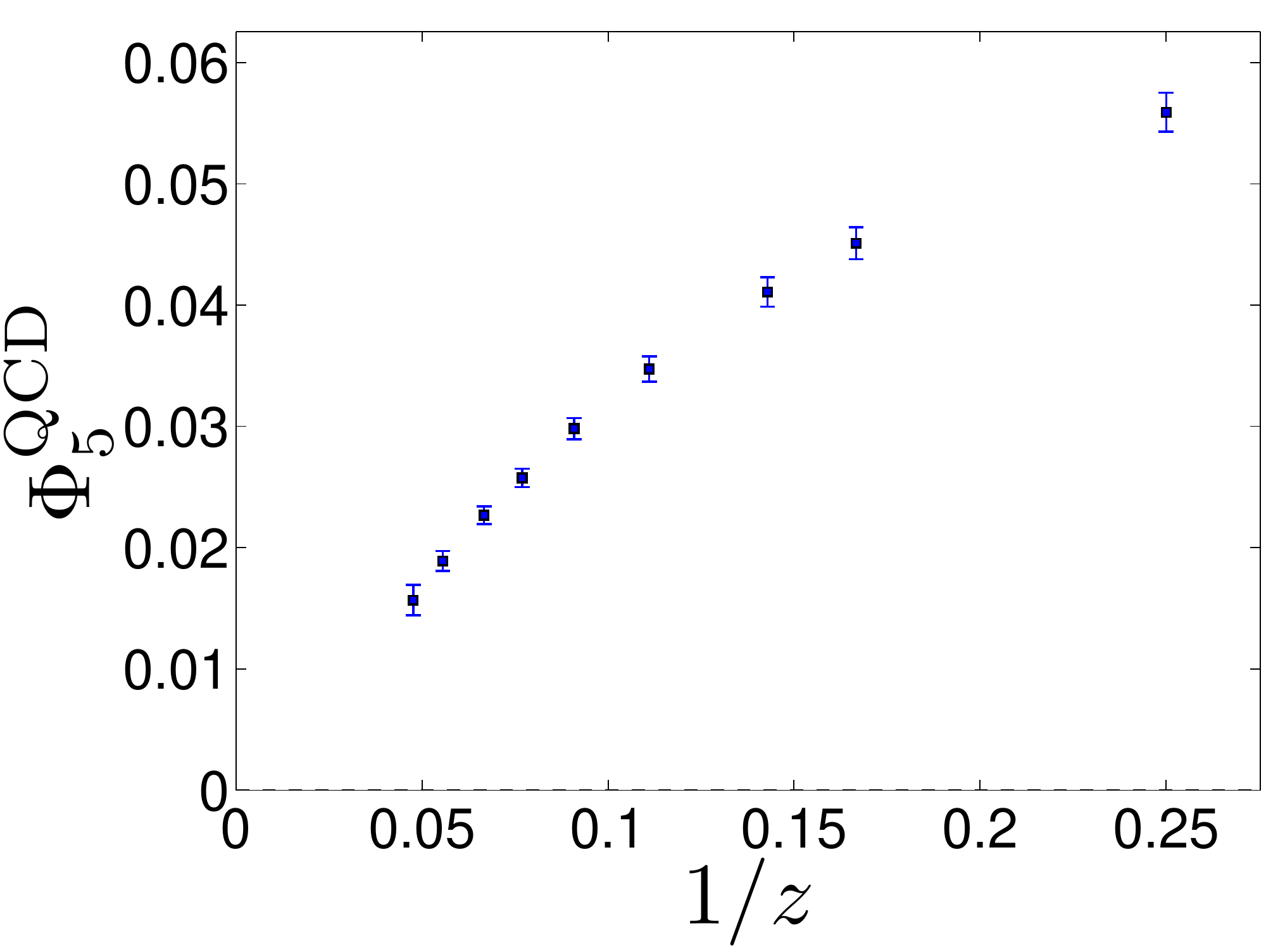}
\end{center}
\vspace{-2em}
\caption{\footnotesize
Same as Fig.~\ref{fig:Phi341om} for the magnetic contribution $\Phi_5^{\first}(L_1,M,0)=\Phi_5^{\rm QCD}(L_1,M,0)$.}
\label{fig:Phi5}
\end{figure}

Then it is interesting to define the $\minv$ contribution to $\Phi_{3,4,5}$,
\bes
\label{eq:Phi1mL1}
\Phi_i^{\first}(L_1,M,0) = \Phi_i^{\rm QCD}(L_1,M,0) - \eta_i(L_1,0)\;, \qquad i=3,4,5,
\ees
and to study the mass behavior of these quantities, where
$\Phi_{4}^{\first}(L_1,M,0)$ and \linebreak 
$\Phi_{5}^{\first}(L_1,M,0)$ are 
a pure kinetic and magnetic correction, respectively.  
The physical
interpretation of $\Phi_{3}^{\first}(L_1,M,0)$ is more subtle and
involves the $\minv$-correction to the time component of the axial current.  Since
they are expected to vanish like $\minv$ at large $\mh$, any strong
deviation from a linear behavior could be interpreted as a contribution from
higher-order corrections of the effective theory%
\footnote{%
Some deviations from linearity are expected since a $\Or(\minv)$-behavior
always contains logarithmic modi\-fications due to the renormalization
of the effective theory.}.
One can see from~\fig{fig:Phi341om} and~\fig{fig:Phi5} that our results for the
heaviest masses are compatible with the expected (linear) leading order
behavior in $1/z$.

\subsection{Matching in $L_1$}
\label{s:3.3}
With the set of simulations described in the previous section, we have computed
$\eta(L_1,a)$ and  $\phimat(L_1,a)$.  Thus the
HQET parameters $\tilde \omega(M,a)$ can be obtained from eq.~\eqref{eq:omega_tilde}, viz.
\begin{equation}
\tilde\omega(M,a) = 
\phimat^{-1}(L_1,a) 
\Big(
\Phi^{\rm QCD}(L_1,M,0) -  \phistat(L_1,a)
\Big)\;.
\notag
\end{equation}
However, in practice we split this equation in the following way
\bes
\tilde\omega_i(M,a) 
&=& \sum_{j=1}^2
\phimat^{-1} _{ij}(L_1,a)
\Big(
\Phi_j^{\rm QCD}(L_1,M,0) -  \phistat_j(L_1,a) \Big) 
\nonumber \\
\label{eq:omegatildesplit}
&+& \sum_{j=3}^5
\phimat^{-1}_{ij}(L_1,a)
\Big(
\Phi_j^{\first}(L_1,M,0) \Big) \;,
\ees
where we have used eq. \eqref{eq:Phi1mL1}.  
Whether one uses \eq{eq:omega_tilde} or \eq{eq:omegatildesplit} to define $\tilde\omega(M,a)$
only affects the way we treat the lattice
artefacts, but one expects a better precision in the second case.  The reason
is that the quantities $\eta_3$ and $\eta_4$ are by far the dominant part of
$\Phi_3$ and $\Phi_4$ (and $\Phi_5$ vanishes in the static approximation). 
As explained in the previous subsection, they are static quantities which extrapolate to the continuum
with $\Or(a^2)$ corrections. 
On the contrary, $\eta_1$ and $\eta_2$ are divergent and have to be kept in the combination as in
eq.~\eqref{eq:omegatildesplit}. They are extrapolated linearly in $a/L_1$.

At this point we would like to comment about the separation of the different
orders in the effective theory.  The situation for the first two observables is
different from the others, since $\eta_1$ and $\eta_2$ do not have a 
continuum limit. However, the whole matching procedure can be
carried out at static order as well. In that case,
$\tilde\omega_3=\cah{1}$ is just the improvement coefficient $a\castat$ which is
approximated by perturbation theory. Moreover, the parameters $\tilde\omega_4$
and $\tilde\omega_5$ are of order $\minv$, and therefore are set to zero.
This setup defines the static approximation of the first two observables,
$\Phi_1^{\rm stat}$ and $\Phi_2^{\rm stat}$,  with $\phimat=\diag(L_1,1)$. 
Performing the matching only for these observables (instead of $\Phi_i^{\rm HQET}$ 
with $i=1,\ldots,5$) allows us to determine the two parameters 
$\mhbarestat$ and $\lnzastat$ at static order.

\subsection{Evolution to a larger volume $L=L_2$}
\label{s:s3.4}
We then consider a set of simulations of HQET in which we use the same parameters as
in the previous step (HQET in volume $L_1$), but where we double the number of
points in each space-time direction. There we compute the quantities
$\phimat(L_2,a)$ and $\phistat(L_2,a)$ with $L_2=2L_1$.  We now have all the
ingredients to compute $\Phi(L_2,M,0)$ using~\eq{eq:Phi_L2}, but for the reason
given above, we use a slightly different version:
\bes
\label{eq:Phi_L2_split}
\Phi_i^{\rm HQET}(L_2,M,0)
= 
\left\{
\begin{array}{l r}
\displaystyle
\lim_{a\to0}
\Big[
\phistat_i(L_2,a) + \sum_{j=1}^5 \phimat_{ij}(L_2,a)\, \tilde\omega_j(M,a)
\Big]\,, &i=1,2  \;,\\[2.0ex]
\displaystyle
\phistat_i(L_2,0) + 
\lim_{a\to0}
\Big[ 
\sum_{j=1}^5 \phimat_{ij}(L_2,a)\, \tilde\omega_j(M,a)
\Big]\,, & i\ge3 \;,
\end{array}
\right.
\ees
Again,
the continuum extrapolations of $\phistat_3$ and $\phistat_4$ are done linearly
in $(a/L_2)^2$, while $\eta_5=0$. All other extrapolations are done 
linearly in $a/L_2$ (see the discussions in Sect.~\ref{s:s3.2} and
Sect.~\ref{s:3.3}).

\begin{figure}[t]
\begin{center}
\begin{tabular}{cc}
\includegraphics[width=0.45\textwidth]{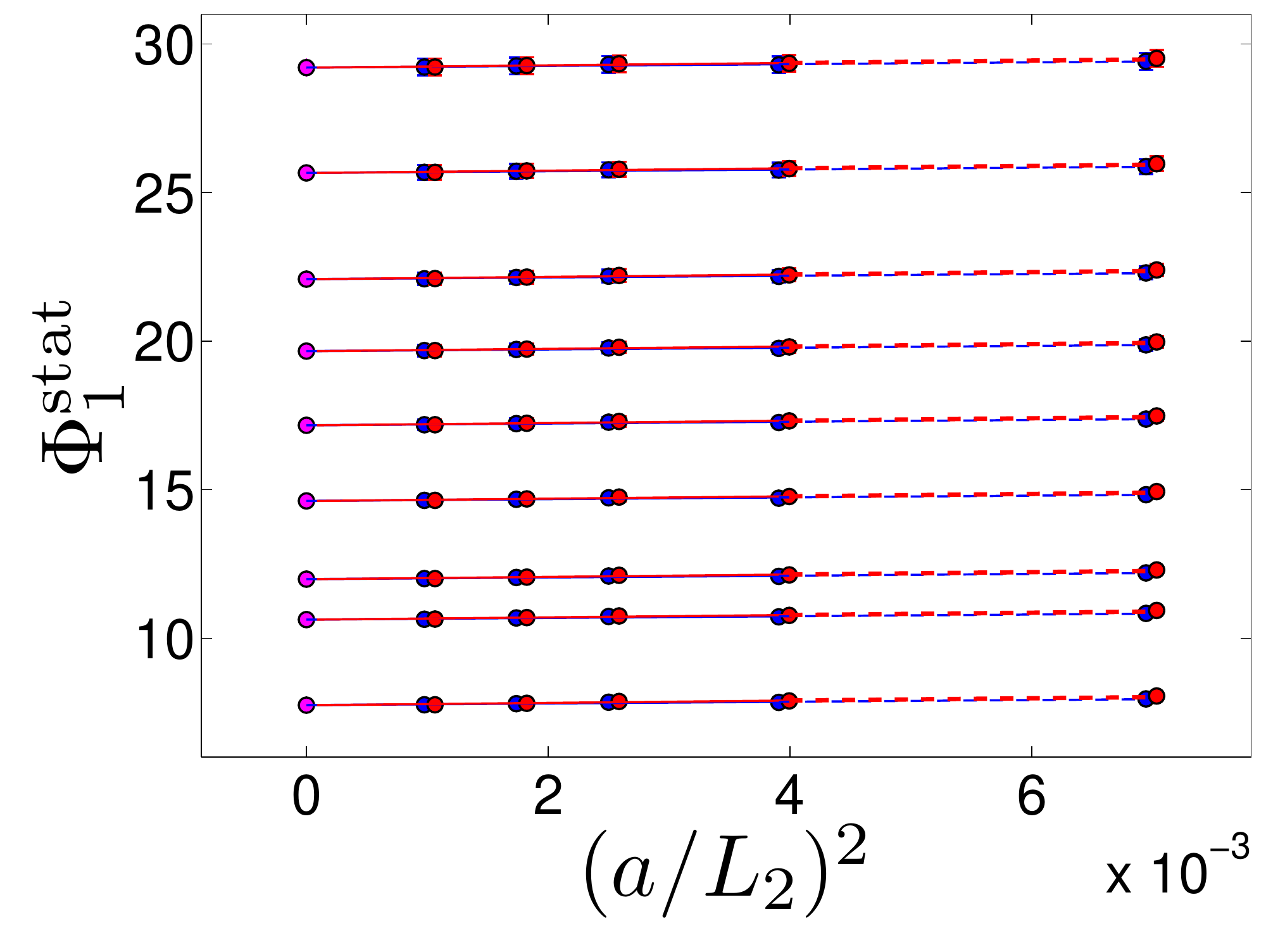}&
\quad
\includegraphics[width=0.45\textwidth]{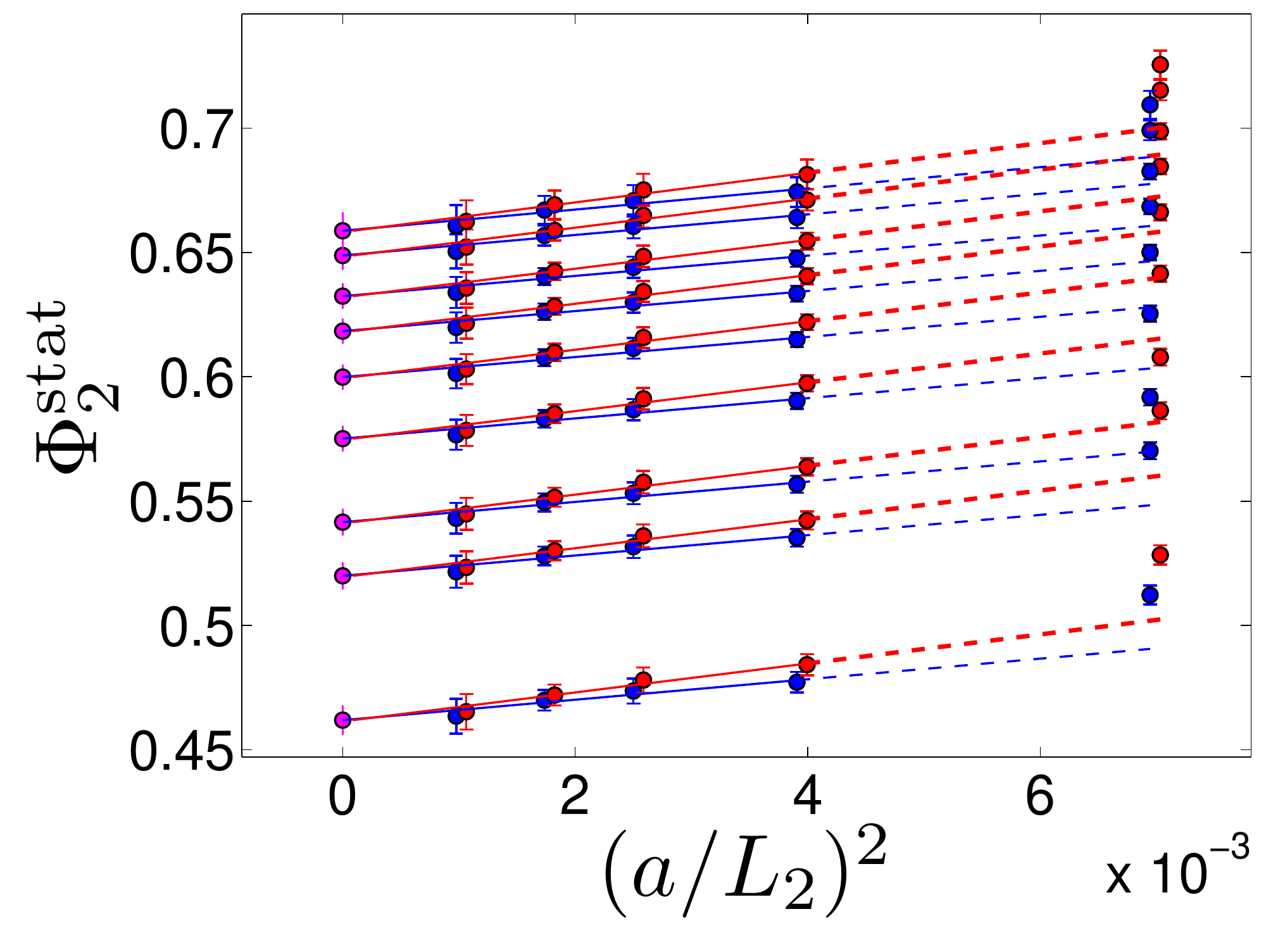}
\end{tabular}
\end{center}
\vspace{-2em}
\caption{\footnotesize
Continuum extrapolation of $\Phi_1(L_2,M,a)$ and $\Phi_2(L_2,M,a)$ 
in the case where the matching is done in the static 
approximation. We show the results for the nine different heavy quark masses
and the two discretizations HYP1,2. The conventions are the same as in \fig{fig:PhiQCD}.}
\label{fig:PhiIstat12}
\end{figure}

\begin{figure}[t]
\begin{center}
\begin{tabular}{cc}
\includegraphics[width=0.45\textwidth]{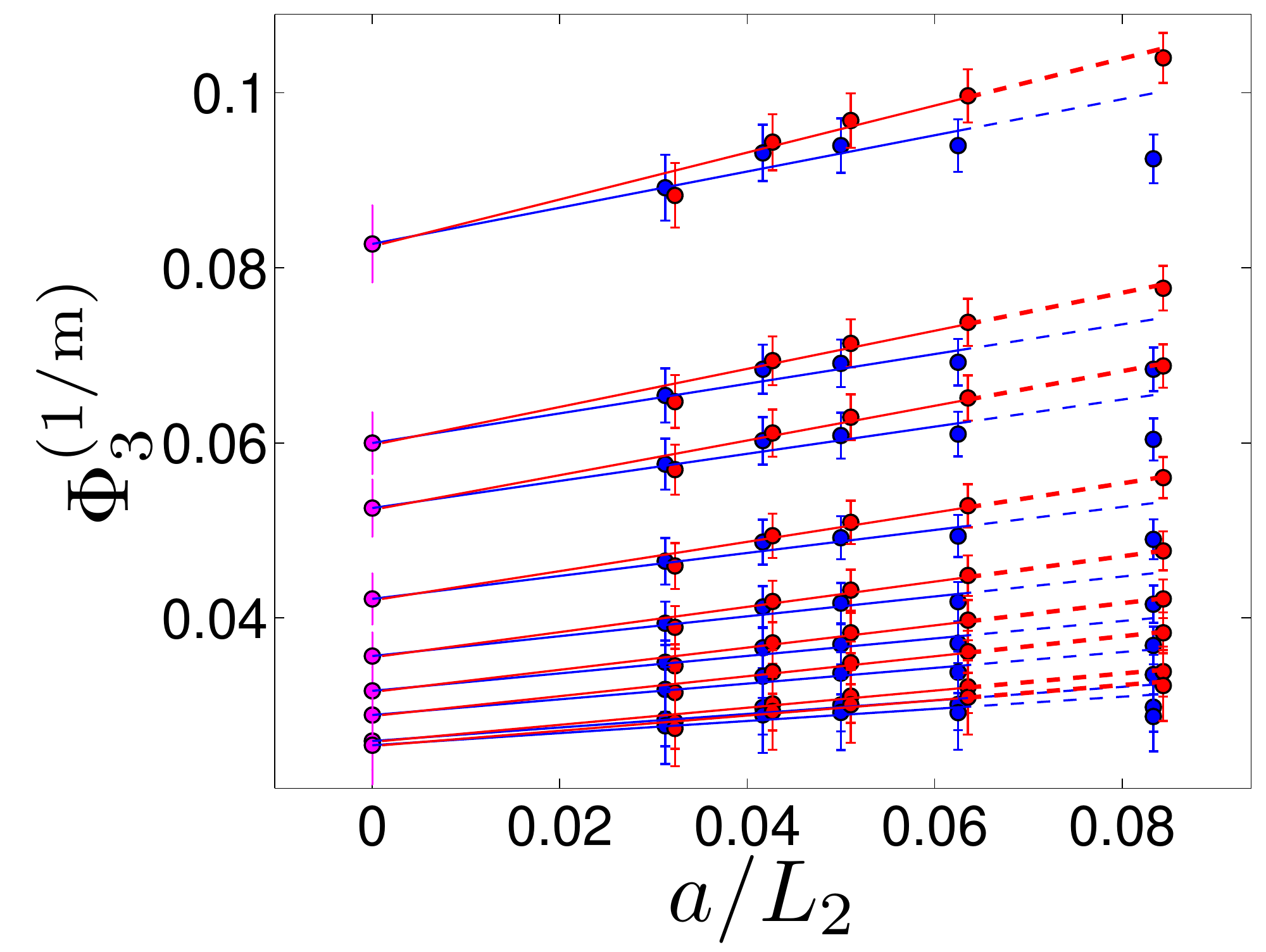}&
\quad
\includegraphics[width=0.45\textwidth]{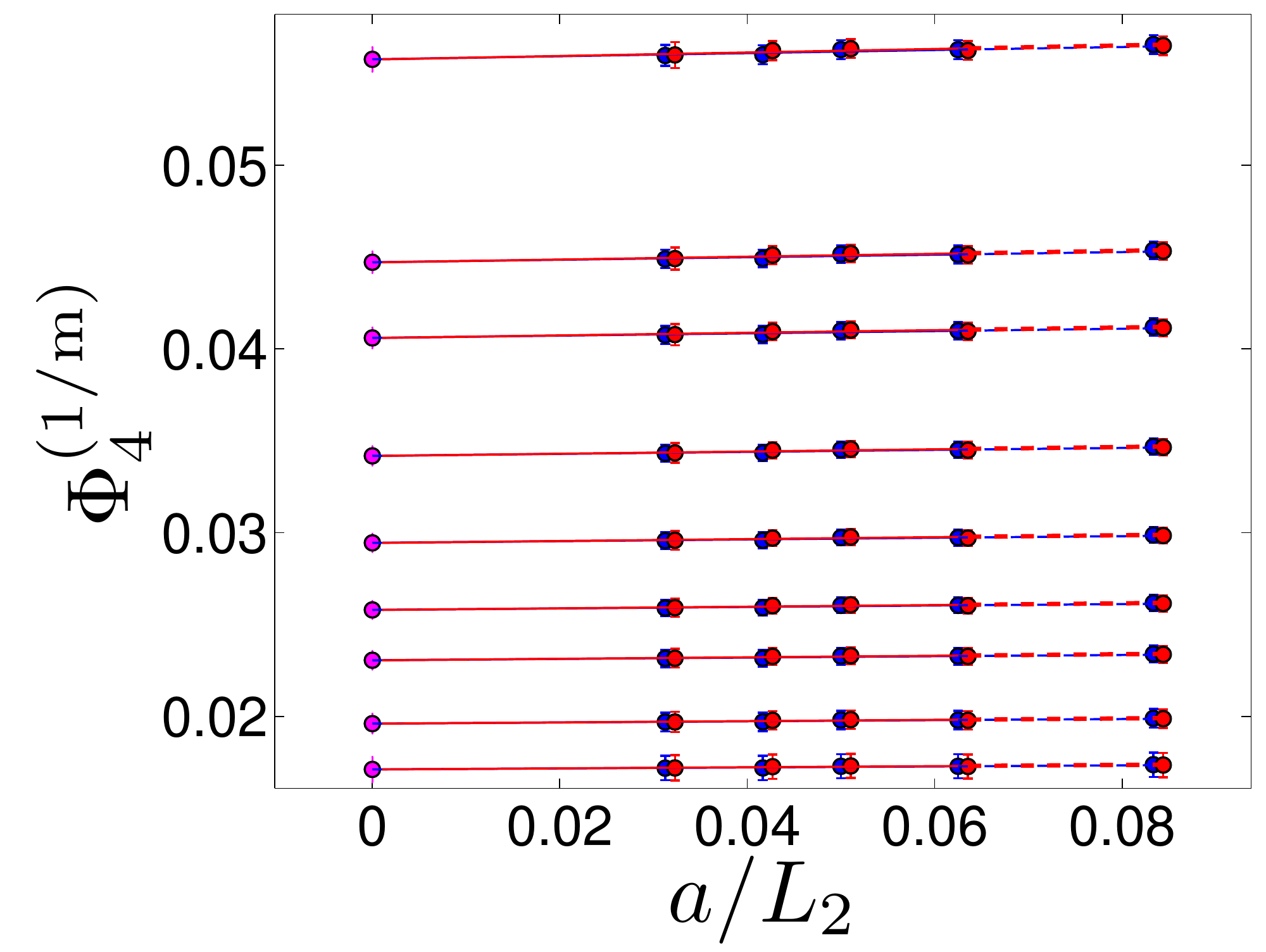}
\end{tabular}
\end{center}
\vspace{-2em}
\caption{\footnotesize Same as~\fig{fig:PhiIstat12} for the $\minv$ contributions  $\Phi^{\first}_3(L_2,M,a)$ 
and $\Phi^{\first}_4(L_2,M,a)$. }
\label{fig:Phi34HQET}
\end{figure}

\begin{figure}[t]
\begin{center}
\includegraphics[width=0.45\textwidth]{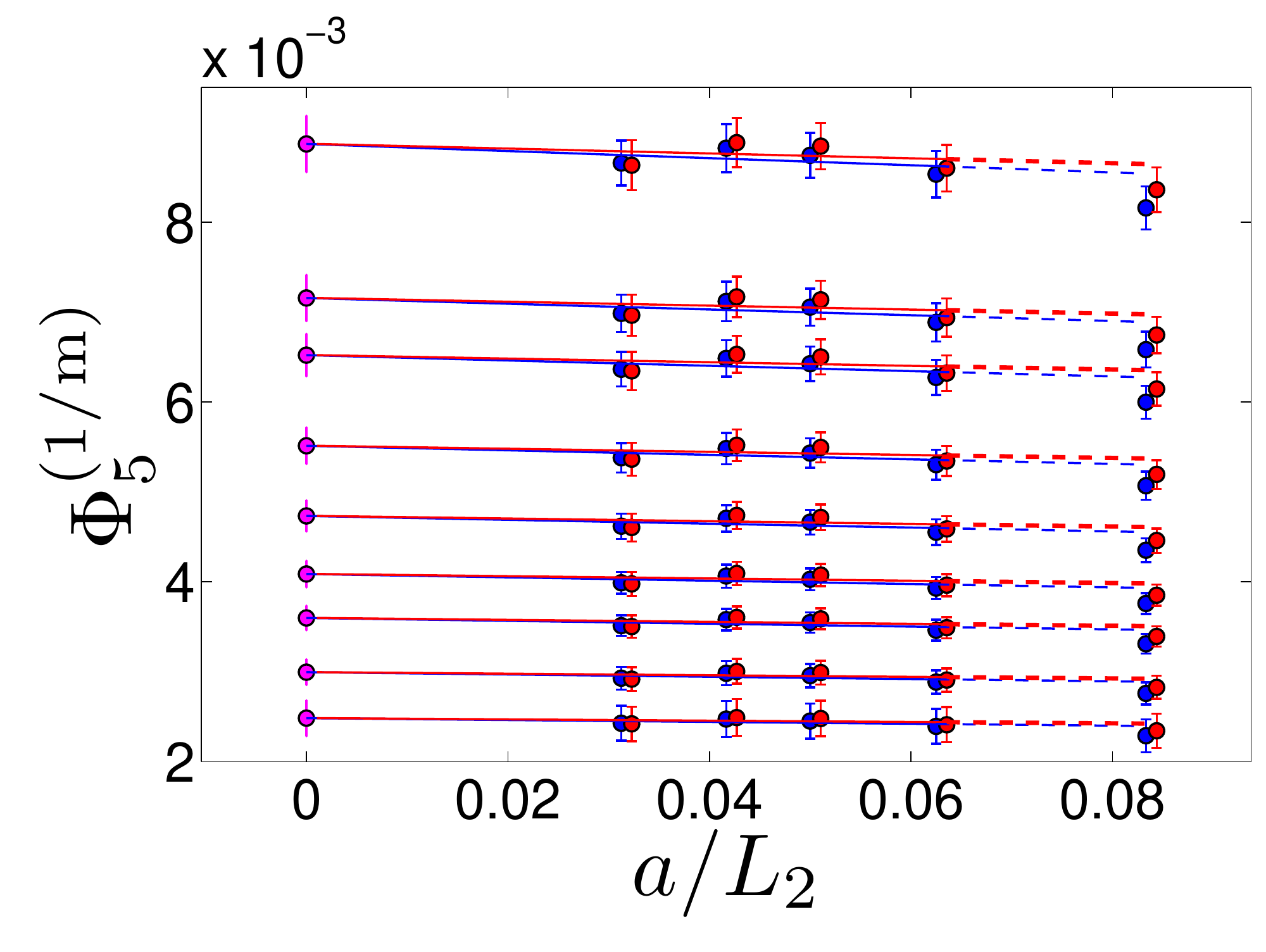}
\end{center}
\vspace{-2.5em}
\caption{\footnotesize
Same as~\fig{fig:PhiIstat12} for $\Phi^{\first}_5(L_2,M,a)$.}
\label{fig:Phi5HQET}
\end{figure}

At the static order, the matching procedure at $L_1$ and the evolution to $L_2$ 
can be done in the same way as with the five-component vector $\Phi$. 
The result defines the quantities $\Phi_i^{\rm stat}(L_2,M,0)$ for $i=1,2$,
and their continuum extrapolation is shown in \fig{fig:PhiIstat12}.
When the matching is performed at the next-to-leading order, we have
$\Phi_i^{\rm stat}(L_2,M,0) =  \eta_i(L_2,0)$ for $i=3,4,5$ and
define the $\minv$-contributions as
\bes
\label{eq:Phi1mL2}
\Phi_i^{\first}(L_2,M,0) &=& \Phi_i^{\rm HQET}(L_2,M,0) - \Phi_i^{\rm stat}(L_2,M,0) \;.
\ees
Their continuum extrapolations are shown in \fig{fig:Phi34HQET} 
and \fig{fig:Phi5HQET}.

The parameters $\omega(M,a)$ are computed 
by the relation analogous to eq.~({\ref{eq:omegatildesplit})
in the volume $L_2$
\bes
\omega_i(M,a) 
&=& \sum_{j=1}^2
\phimat^{-1}_{ij}(L_2,a) 
\Big(
\Phi_j^{\rm HQET}(L_2,M,0) -  \phistat_j(L_2,a) \Big) 
\nonumber \\
\label{eq:omegasplit}
&+& \sum_{j=3}^5
\phimat^{-1}_{ij}(L_2,a) 
\Big(
\Phi_j^{\first}(L_2,M,0) \Big) \;.
\ees

In~\eq{eq:Phi_L2_split} we have chosen to write the evolution of the
observables to the volume $L_2$ in terms of the HQET couplings $\tilde\omega_i$.
Equivalently we can introduce a matrix of step scaling functions.  In order to
do that, one substitutes in \eq{eq:Phi_L2_split} the $\tilde\omega_i$ by the matching
\eq{eq:omega_tilde}.  One obtains an equation of the following form:
\bes
\Phi_i(L_2,M,0) 
&=& 
D_i\, \Phi_i(L_1,M,0) + \lim_{a/L_1\to0}\,\widehat\Sigma_{i}(L_1,a) 
\nonumber \\
&+& 
\lim_{a/L_1\to0} \sum_{j=3}^5\, \Sigma_{ij}(L_1,a) \Phi_j^{\first}(L_1,M,0) \;,\quad i=1,2\,,   
\\
\Phi_i^{\first}(L_2,M,0) 
&=& 
\lim_{a/L_1\to0} \sum_{j=3}^5\Sigma_{ij}(L_1,M,a)  \Phi_j^{\first}(L_1,M,0) \;,
\quad i=3,4,5\,.
\ees
The explicit definitions of $D, \Sigma$ and $\widehat\Sigma$ can be found
in~\cite{Blossier:2010jk}. 
We have implemented the tree-level improvement of these step-scaling
functions in order to obtain a smoother approach to the continuum limit.
Our results are obtained in this way, but tree-level improvement
actually only has a small influence on our results after extrapolation.

\section{HQET parameters to be used in large volume simulations}
\label{s:s4}
As we have already mentioned in the text and explained in detail in the
appendices, the HQET parameters are obtained at five values of the bare
coupling $g_0^2$, which are such that the renormalized coupling $\gbar^2(L_2/4)$
is kept constant for the different ensembles.  This is done by setting $\beta$ to some
precise values: $\beta = 5.2638, 5.4689, 5.619, 5.758, 5.9631$.  In order to be
able to use the HQET parameters in large volume simulations -- and make phenomenological
predictions --  we interpolate
(or, in one case, we extrapolate) them to $\beta=5.2, 5.3, 5.5, 5.7$, using a
quadratic polynomial in $\beta$.  We show this interpolation/extrapolation
in~\fig{fig:mbare_interp}.  The results are reported in \Tab{table_hqet_param}
for the action HYP2. 
All errors quoted in this table are statistical (obtained with a standard 
jackknife procedure), including errors coming from the QCD renormalization constants
at finite lattice spacing.
There is still an overall relative error contribution of about 0.9\% from the 
quark mass renormalization in QCD, i.e., the factor $h$ in eq.~\eqref{eqn:h-L0} 
of \ref{a:tuning}, which relates the RGI mass to the SF running mass in the 
continuum limit.
Since in practice this error will only become relevant when the $z$ dependence 
of the HQET parameters is actually applied to interpolate to the B-meson scale 
and to extract physical quantities, it is enough to include it then. 
The complete set of results can be found on our website \url{http://www-zeuthen.desy.de/alpha/}, 
together with the relevant error-correlation matrices.  From preliminary
studies~\cite{Blossier:2010vj,Blossier:2011dk}, we know that the physical
b-quark mass corresponds approximately to $z=13$.  For this value of $z$
we display our results for $a\mhbare$ rescaled by $L_1/a$ for each value of the
lattice spacing in~\fig{fig:mbare_div}.
The parameter $\mhbare$ absorbs the power divergences present in the binding energy
of an heavy-light meson in HQET: a $1/a$ divergence at the static order, and a
$1/a^2$ divergence at the $\minv$ order.  As one can see from the figure,
$\mhbare$ is dominated by these power divergences.  Clearly, in order to
guarantee the existence of the continuum limit, these divergences have to be
removed non-perturbatively, which is one of the benefits of our approach. 

\TABLE{%
 \hspace{-1.cm} 
\small
\renewcommand{\arraystretch}{1.25}
 \begin{tabular}{cccccc}\toprule
  & $\beta$          &  5.7 &  5.5 &  5.3 & 5.2 \\ \midrule 
                    & $z=11$ & $0.601(8)$   & $ 0.833(10)$  &  $1.129(13)$ &  $1.302(15)$ \\    
 $a\mhbare^\stat $  & $z=13$ & $0.713(9)$   & $ 0.982(11)$  &  $1.327(15)$ &  $1.527(17)$ \\ 
                    & $z=15$ & $0.821(10)$  & $ 1.127(13)$  &  $1.518(17)$ &  $1.746(19)$ \\\cmidrule(lr){1-6} 
                    & $z=11$ & $-0.209(4)$  &  $-0.201(5)$  &  $-0.197(5)$ &  $-0.197(5)$ \\                   
 $\lnzastat$        & $z=13$ & $-0.191(4)$  &  $-0.183(5)$  &  $-0.179(5)$ &  $-0.178(5)$ \\                   
                    & $z=15$ & $-0.176(4)$  &  $-0.169(5)$  &  $-0.165(5)$ &  $-0.164(5)$ \\\midrule

                    & $z=11$ & $ 0.058(12)$ &  $ 0.377(13)$ &  $ 0.740(21)$ &  $ 0.937(17)$\\ 
 $a\mhbare$         & $z=13$ & $ 0.236(12)$ &  $ 0.582(14)$ &  $ 0.985(17)$ &  $ 1.207(18)$\\ 
                    & $z=15$ & $ 0.395(13)$ &  $ 0.769(15)$ &  $ 1.212(18)$ &  $ 1.459(20)$\\\cmidrule(lr){1-6}
                    & $z=11$ & $-0.212(42)$ &  $-0.194(37)$ &  $-0.173(34)$ &  $-0.162(33)$\\ 
 $\lnzahqet$        & $z=13$ & $-0.181(40)$ &  $-0.166(36)$ &  $-0.148(32)$ &  $-0.139(31)$\\ 
                    & $z=15$ & $-0.155(41)$ &  $-0.142(36)$ &  $-0.127(32)$ &  $-0.119(31)$\\\cmidrule(lr){1-6}
                    & $z=11$ & $-1.00(14)$  &  $-0.75(12)$  &  $-0.63(10)$  &  $-0.60(9)$\\
 $\cahqet/a$        & $z=13$ & $-0.89(14)$  &  $-0.68(12)$  &  $-0.56(10)$  &  $-0.54(9)$\\ 
                    & $z=15$ & $-0.83(15)$  &  $-0.63(12)$  &  $-0.52(10)$  &  $-0.50(9)$\\\cmidrule(lr){1-6}
                    & $z=11$ & $ 0.778(13)$ &  $ 0.608(11)$ &  $ 0.485(8)$ &  $ 0.441(7)$ \\ 
 $\omegakin/a$      & $z=13$ & $ 0.682(13)$ &  $ 0.533(10)$ &  $ 0.425(8)$ &  $ 0.386(7)$ \\ 
                    & $z=15$ & $ 0.609(13)$ &  $ 0.476(11)$ &  $ 0.380(8)$ &  $ 0.345(7)$ \\\cmidrule(lr){1-6}
                    & $z=11$ & $ 1.626(59)$ &  $ 1.284(49)$ &  $ 1.041(39)$ &  $ 0.956(35)$ \\
 $\omegaspin/a$     & $z=13$ & $ 1.404(51)$ &  $ 1.109(42)$ &  $ 0.899(34)$ &  $ 0.825(30)$ \\
                    & $z=15$ & $ 1.236(47)$ &  $ 0.976(39)$ &  $ 0.791(31)$ &  $ 0.727(28)$ \\
 \bottomrule
 \end{tabular} 
 \caption{\footnotesize HQET parameters as a function of the bare coupling
 for the action HYP2 at  $z=11,13,15$. The central value $z=13$ is close 
 to the physical b-quark mass. The first two parameters, $\mhbare^\stat$ and $\lnzastat$, result 
 from  the matching at static order, while the remaining entries are the
 parameter set resulting from the matching of HQET at next-to-leading order.}
 \label{table_hqet_param}

}
%
\begin{figure}
\begin{center}
\begin{tabular}{cc}
\includegraphics[width=0.49\textwidth]{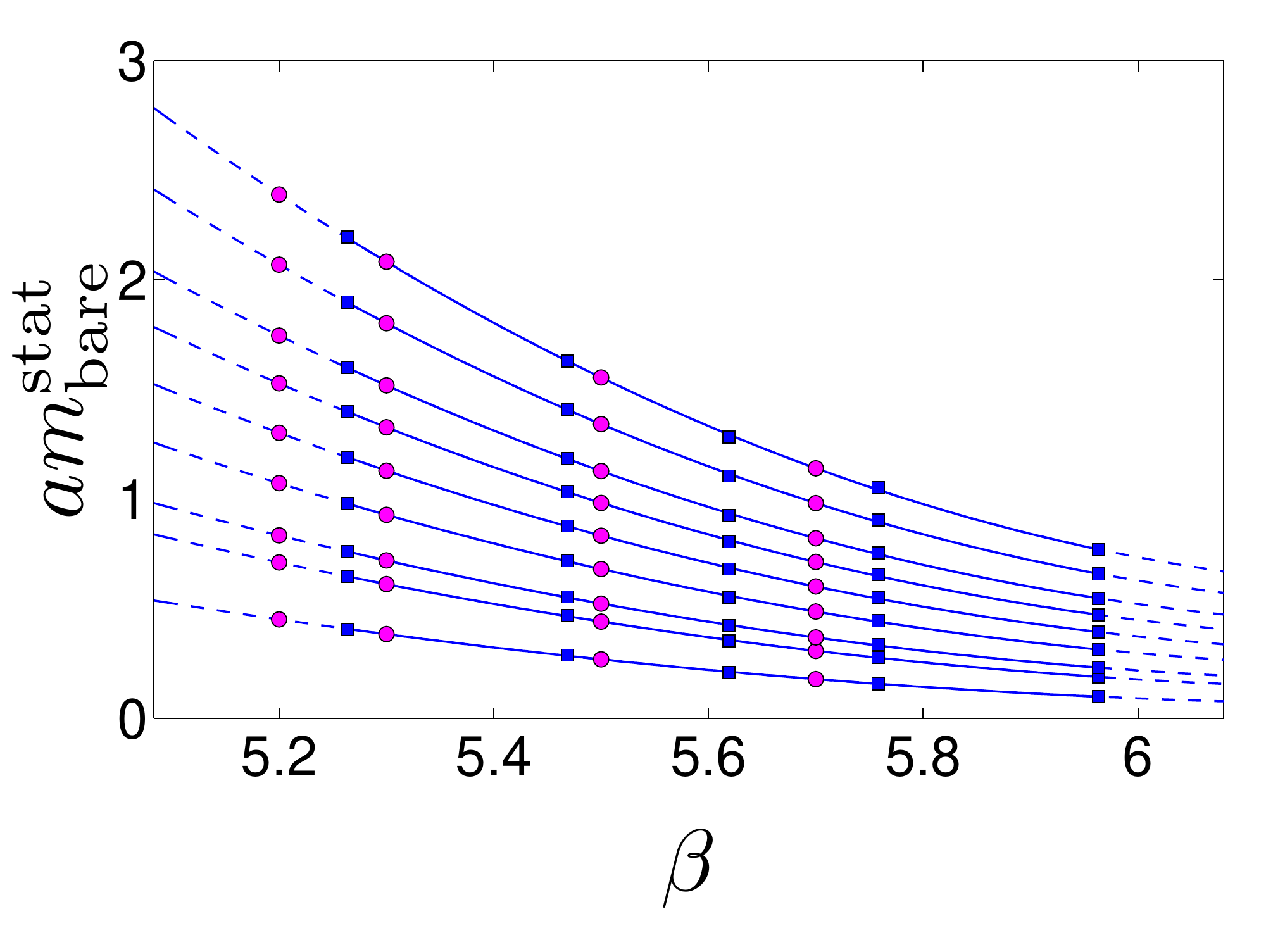}&
\includegraphics[width=0.49\textwidth]{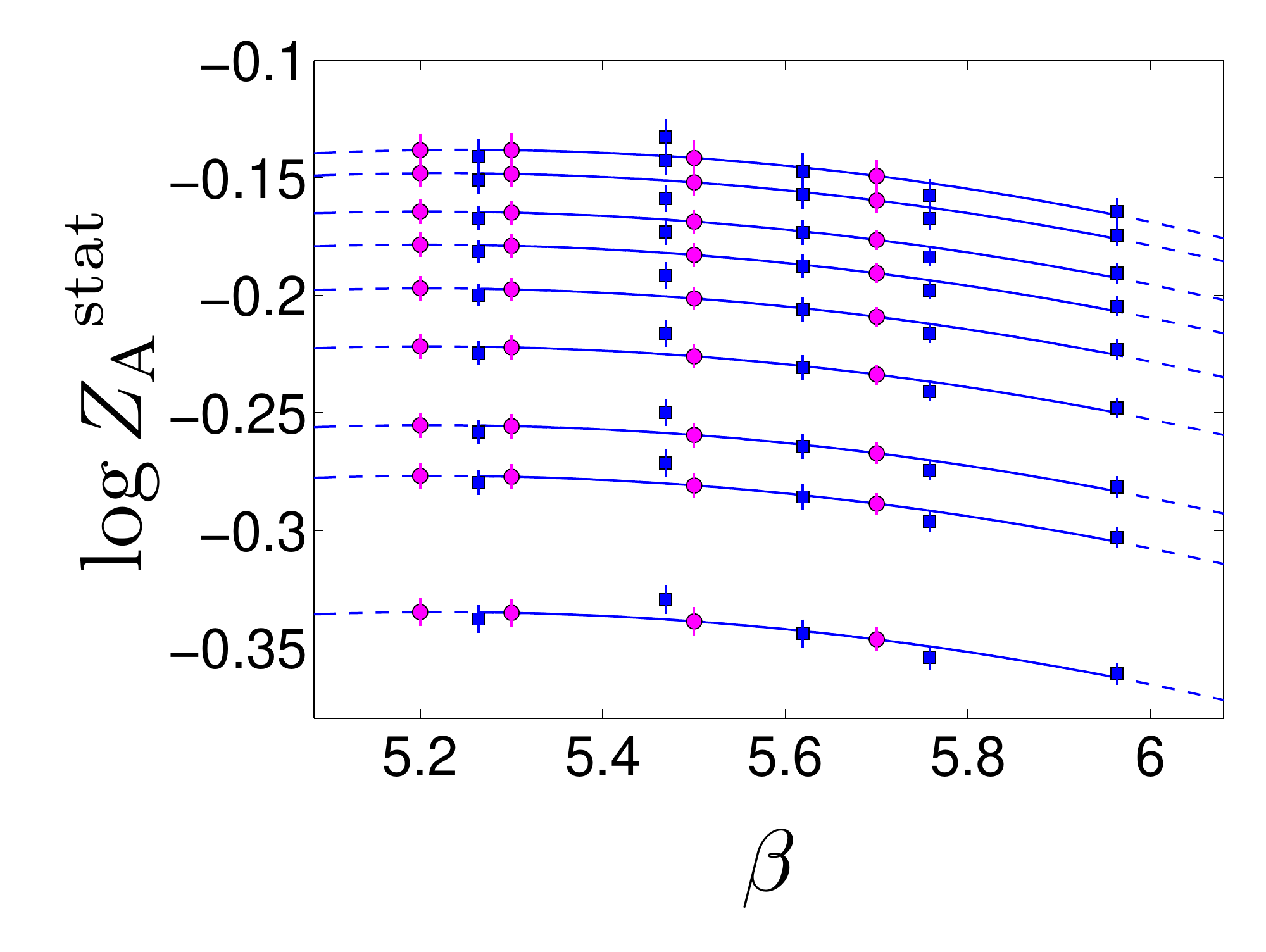}\\[-1em]
\includegraphics[width=0.49\textwidth]{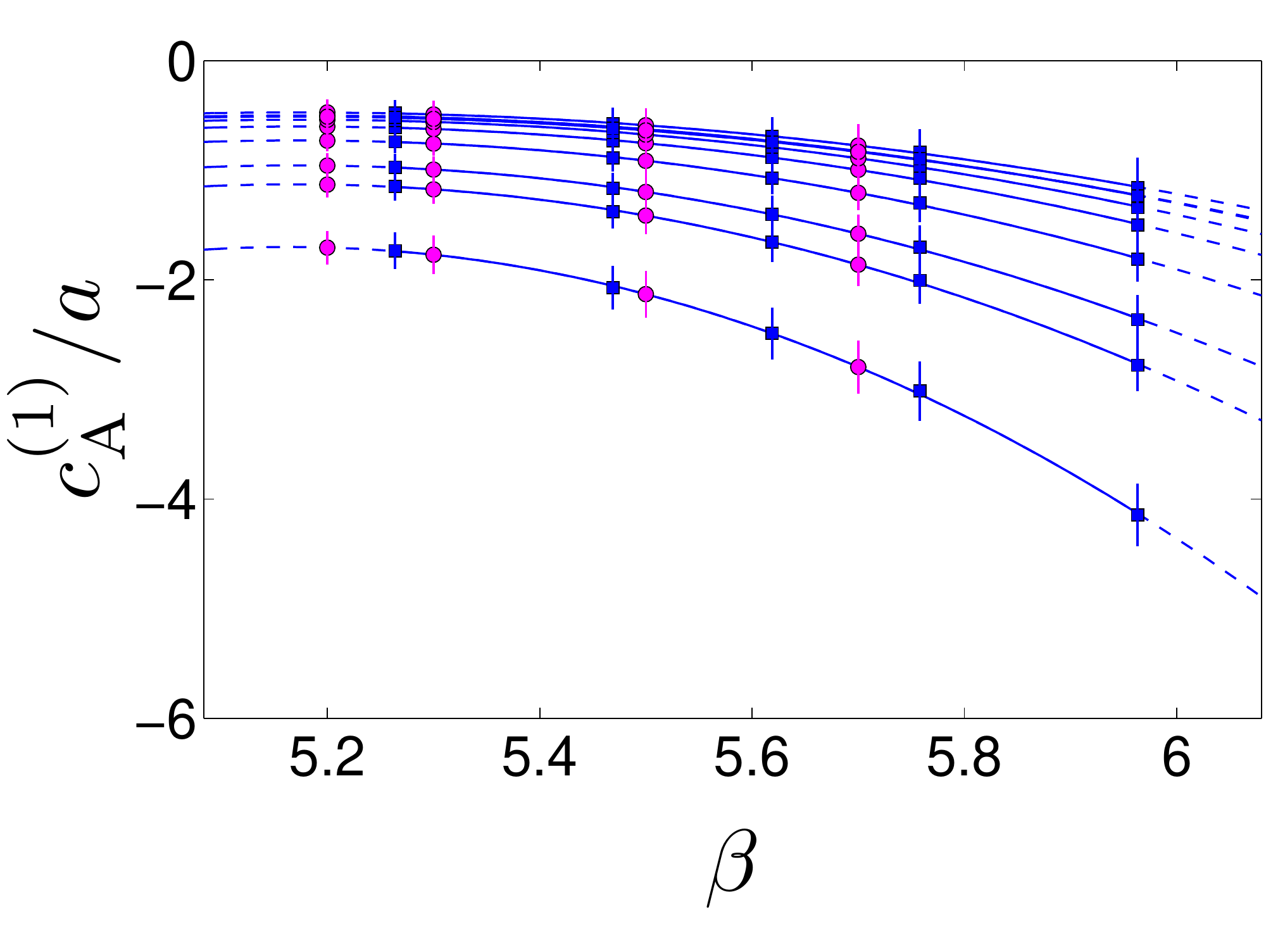}&
\includegraphics[width=0.49\textwidth]{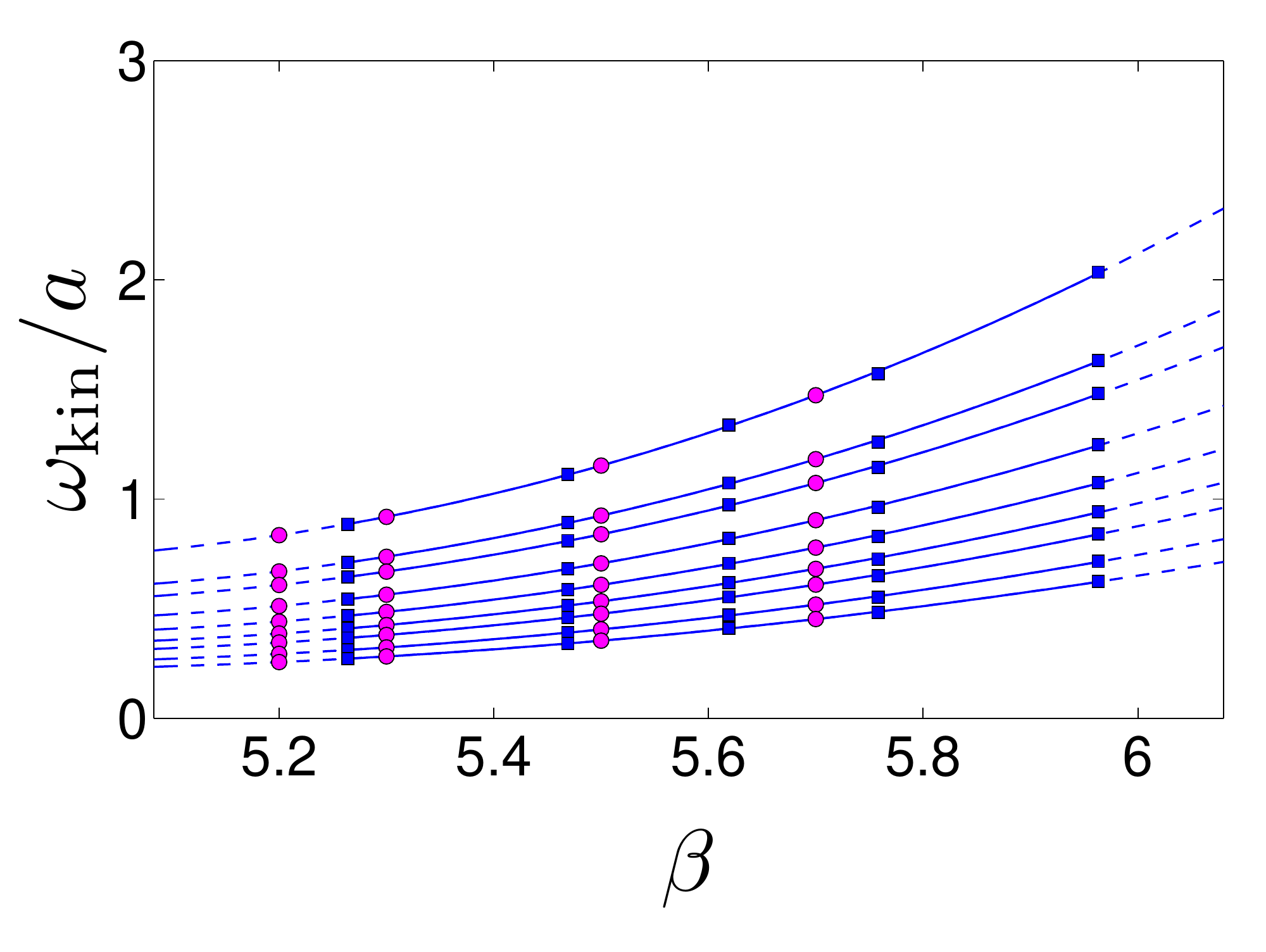}\\[-1em]
\end{tabular}
\includegraphics[width=0.49\textwidth]{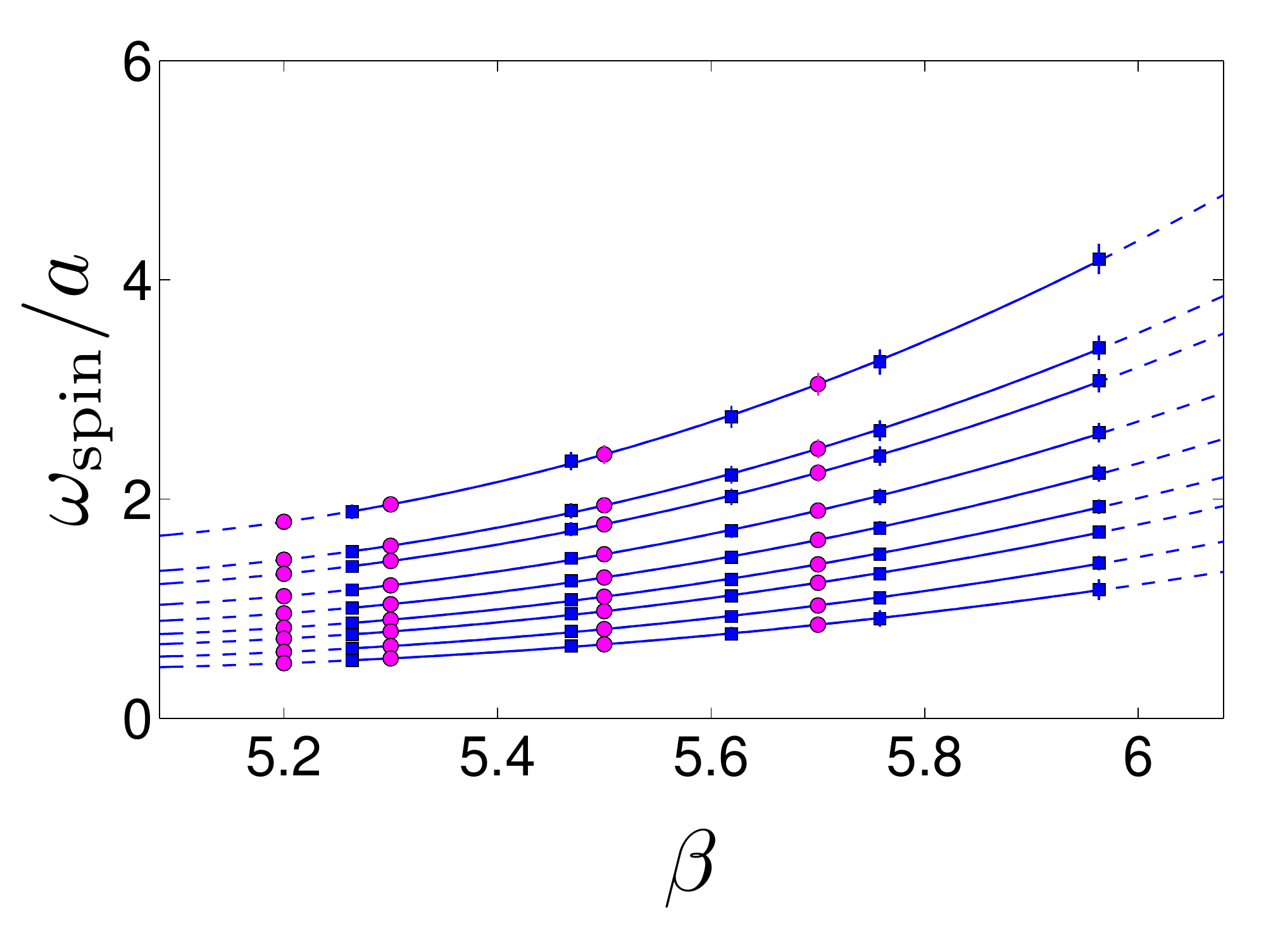}
\vspace{-1.5em}
\caption{\footnotesize
Quadratic fits of HQET parameters in $L_2$ as a function of $\beta$. The blue points
are the results of the numerical simulations and the magenta points are 
the interpolated (extrapolated) ones. In this plot we show the results for 
the HYP2 discretization and for the nine different values of the heavy quark mass. 
}
\label{fig:mbare_interp}
\end{center}
\end{figure}

\begin{figure}
\begin{center}
\begin{tabular}{cc}
\includegraphics[width=0.45\textwidth]{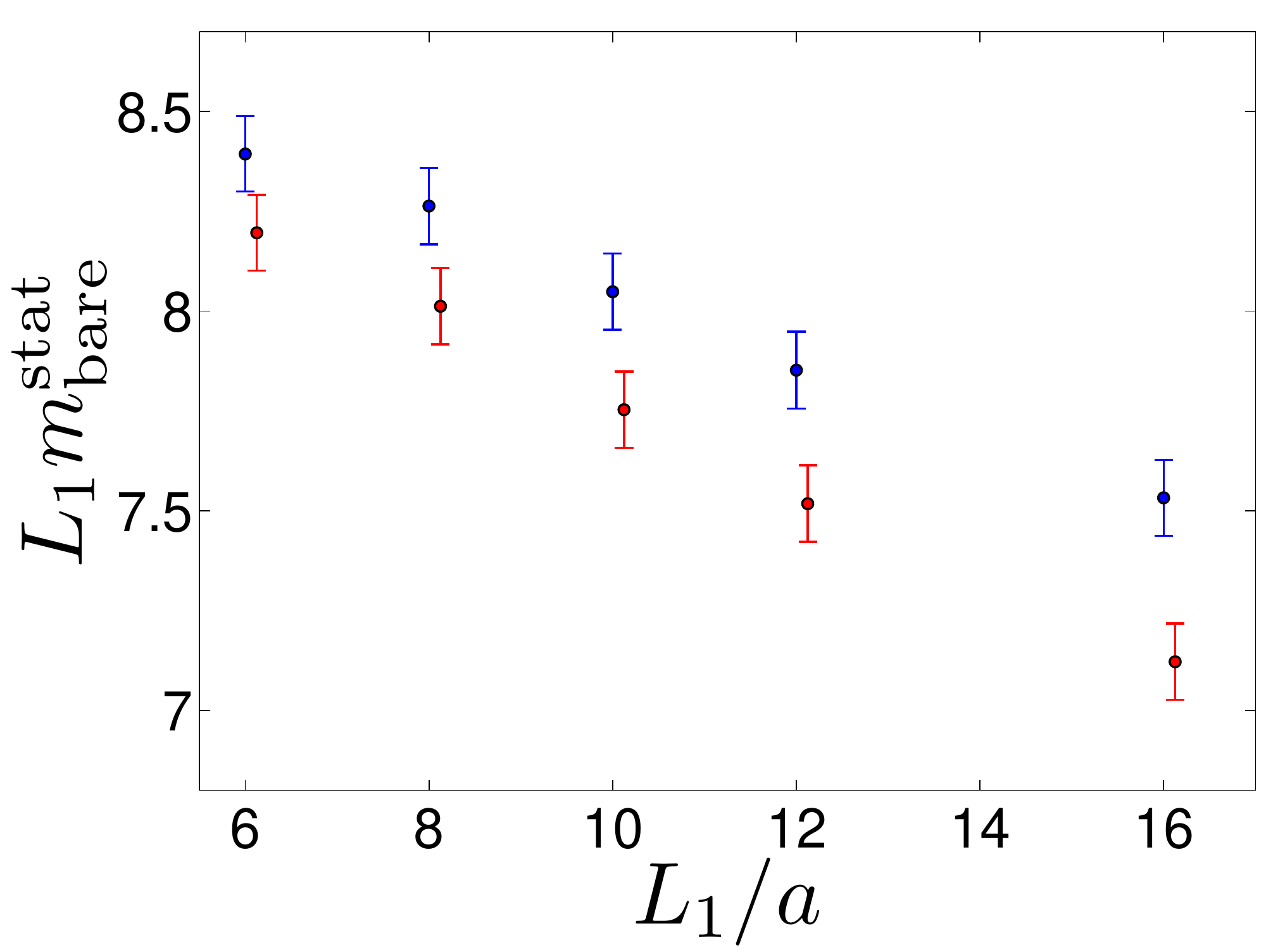}&
\includegraphics[width=0.45\textwidth]{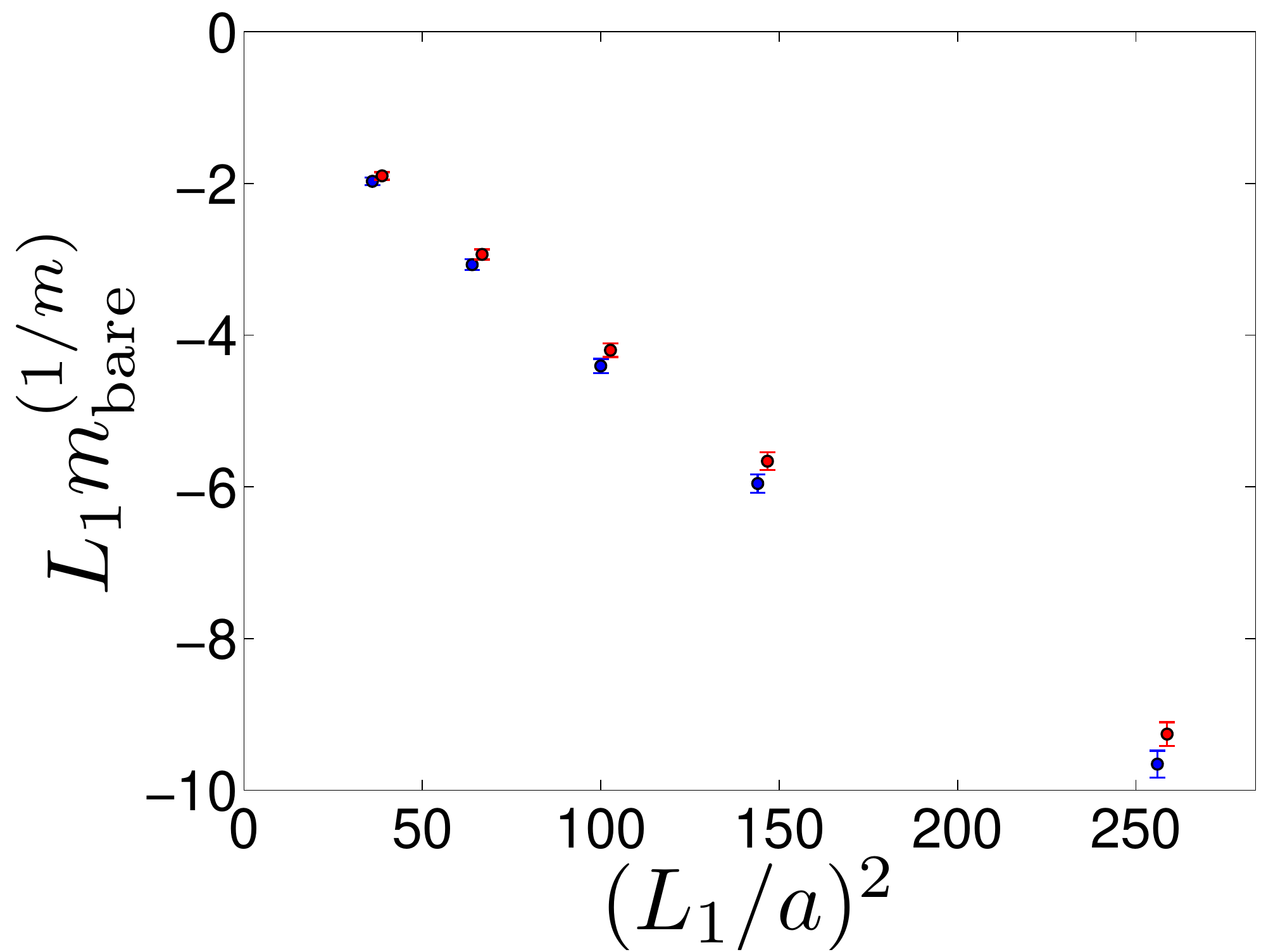}\\
\end{tabular}
\vspace{-1em}
\caption{\footnotesize
The bare quark mass at the static order and its $\minv$ correction, in units of $L_1$,
for $z=13$ and the two discretizations HYP1,2. 
These quantities are expected to absorb the power divergences $(1/a)^n$, 
with $n=1$ for the static order and $n=2$
for the $\minv$ correction.
}
\label{fig:mbare_div}
\end{center}
\end{figure}

Finally, we would like to comment on the size of $\minv^2$ terms. In our setup,
the fermion  fields are periodic (in space) up to a phase.  As a consequence,
the static quantities depend on the choice of one angle, that we call
$\theta_0$, and the quantities computed at the next-to-leading order of HQET
depend on three angles: $\theta_0,\theta_1,\theta_2$. So far in this work we
have considered only what we call the standard choice of $\theta$
angles~\cite{DellaMorte:2006cb,Blossier:2010jk}, namely
$(\theta_0,\theta_1,\theta_2) = (0.5, 0.5, 1)$.  When we change the values of
$\theta_i$ we change the set of observables, and thereby the matching
conditions. The static quantities computed for different values of $\theta$ are
thus expected to differ by terms of order $\minv$. Once the $\minv$ corrections
are added, the difference should be of order $\minv^2$.  In general we find no
significant $\theta$ dependence for all the HQET parameters $\omega_i$ at the
next-to-leading order of the effective theory, meaning that the $\minv^2$ terms
are not visible within our statistical precision.  As an illustration, we show
the spread of our results for $\lnzastat$ and $\lnzahqet$ in
Table~\ref{table_hqet_param_theta_diff} for the discretization HYP2, 
$\beta=5.3$, and for $z=13$. Due to statistical correlations,
some of the errors on these $\minv^2$ terms
are significantly smaller than our errors for $\lnzastat$ and $\lnzahqet$
themselves, but still no significant $\minv^2$ term is found.
\TABLE{%
\small
\renewcommand{\arraystretch}{1.25}
 \begin{tabular}{ccccc}\toprule
      &  $\Delta\lnzastat$ &\multicolumn{3}{c}{$\Delta\lnzahqet$}\\\cmidrule(lr){2-2}\cmidrule(lr){3-5}
   $(\theta_1,\theta_2)$:       &  &$(0,0.5)$ & $(0.5,1)$ &$(0,1)$\\\midrule
  $\theta_0=0.0$ &  $0.014(2)$  & $-0.046(47)$ &  $-0.001(6)$ &   $-0.012(14)$\\   
  $\theta_0=0.5$ &  $0$         & $-0.048(49)$ &  $ 0 $       &   $-0.012(12)$\\  
  $\theta_0=1.0$ &  $-0.046(2)$ & $-0.055(60)$ &  $ 0.007(7)$ &   $-0.008(18)$\\\bottomrule 
 \end{tabular} 
 \caption{\footnotesize For each combination of $\theta$-angles we compute the 
differences 
 $\Delta\lnzastat=\lnzastat-\lnzastat|_{(\theta_0,\theta_1,\theta_2) = (0.5, 0.5, 1)}$ and 
 $\Delta\lnzahqet=\lnzahqet-\lnzahqet|_{(\theta_0,\theta_1,\theta_2) = (0.5, 0.5, 1)}$ with 
 the HYP2-discretization at $\beta=5.3$ and $z=13$.
 As explained  in the text, different values of $\theta$ correspond  to different 
matching conditions. The small $\theta$-dependence observed at the static order is completely absorbed 
by the $\minv$ corrections, meaning that the $\minv^2$ corrections are not visible within our statistical
errors for that quantity.
 } 
 \label{table_hqet_param_theta_diff}

}

\section{Conclusions}
\label{s:s5}

We have reported on the first unquenched determination of a set of HQET parameters.
While other approaches treat the dependence of the effective parameters on the
renormalized coupling constant in a perturbative fashion, our approach is
entirely non-perturbative in the QCD coupling constant 
and avoids
the difficult-to-estimate uncertainties of perturbation theory
for this case~\cite{Bekavac:2009zc,Sommer:2010ic}\footnote{Ref.~\cite{Bekavac:2009zc}
computes the matching of the static-light currents including $\Or(\alpha_{\rm s}^3)$ 
and comments on a bad behavior of perturbation theory. In Section 3.3.2 of
\cite{Sommer:2010ic}, it is shown that different choices for the renormalization
scale do not really improve the situation. At present no cure seems available
except for a direct non-perturbative matching. 
}.
Furthermore, our matching procedure takes into account
non-perturbatively the power-law divergences of the effective theory which can be 
numerically dangerous if simply addressed 
by relying on perturbation theory. We have shown 
that these divergences get absorbed into the effective parameters of the theory.
The results presented here can be combined with hadronic matrix elements and energies
computed in large volume simulations (some preliminary results have been 
reported in~\cite{Blossier:2010vj,Garron:2011eh,Blossier:2011dk}).
We look forward to presenting our determination of the b-quark mass,
of the B-meson spectrum and of heavy-light decay constants at the $\minv$
order of HQET in the two flavor theory, which will use the parameters computed
in this work.

Concerning the decay constant, the precision of the current matching
is about 0.5\% in the static approximation and 3\% to first order in
$\minv$. The latter puts a mild restriction on the available precision
for the decay constant. It is reassuring to see that in the cases
where we can check for contributions of $\minv^2$ terms explicitly, these 
are considerably below our precision. This is in line with the 
strong hierarchy that we observe between different orders
in the HQET expansion. Around the b-quark mass, the numerical values of 
$\Phi_i(L_2,M,0)$ change from 
$\Phi_1 \approx 20 = \Or(L_2\mbeauty) $ over 
$\Phi_2 \approx 0.6  = \Or(1)$ to $\Phi_{i>2}^{\first} \approx 0.006 - 0.06 = \Or(1/\mbeauty)$, 
as one can see in \fig{fig:PhiIstat12} --
\fig{fig:Phi5HQET}}.

We finally note that our numerical computations have been 
carried out on apeNEXT computers, decommissioned by now. A future application
of our matching strategy with three or more flavors can be expected to reach 
a further improved numerical precision on present or future hardware.

\vspace{0.5cm}
\noindent
{\bf Acknowledgements}\\
We thank our colleagues from the ALPHA Collaboration for many discussions
and in particular Antoine~G\'erardin 
for performing a thorough check of the analysis.
This work is supported by the Deutsche Forschungsgemeinschaft (DFG) in the SFB/TR 09 
and by the European 
community through EU Contract No.~MRTN-CT-2006-035482 (FLAVIAnet).
N.~G. is supported by the STFC grant ST/G000522/1 and 
acknowledges the EU grant 238353 (STRONGnet).
P.~F. and J.~H. acknowledge partial support by the 
DFG under grant \mbox{HE~4517/2-1}. 
N.~T. acknowledges the MIUR (Italy) for partial support under the contract PRIN09.
We thank NIC/DESY and INFN for allocating computer time on the apeNEXT computers 
to this project 
as well as the staff of the computer centers at Zeuthen 
and Rome for their valuable support.

\begin{appendix}
\section{Observables}
\label{a:observables}

For completeness, we remind the reader of the definitions of the observables $\Phi$
introduced in~\cite{Blossier:2010jk},
\bes
\Phi^{\rm QCD} =     \left(L\meffp 
      \,,\; \ln\big[{-\fa}\big/{\sqrt{\fone}}\big]
      \,,\; \ra
      \,,\; \rone
      \,,\; {3\over 4}\ln\big[{\fone}\big/{\kone}\big]
\right)^\mrm{t}\,,
\ees
where the quantities
\bea
\meffp &=& 
-\tilde\partial_0\ln\big[-\fa(x_0,\theta_0)\big]_{(x_0=T/2,\,T=L)}  \;,\\[0.5em]
\ra    &=& \ln\big[\fa(x_0,\theta_1)\big/\fa(x_0,\theta_2)\big]_{(x_0=T/2,\,T=L)}  \;,\\[0.5em]
\rone  &=& {1\over 4}
\ln
\left[
{\fone(\theta_1) \over \fone(\theta_2) } {\kone(\theta_1)^3\over\kone(\theta_2)^3}
\right]_{(T=L/2)}\,,
\eea 
are built from the (renormalized) SF boundary-to-bulk correlation function $\fa$
of the pseudoscalar channel and the boundary-to-boundary correlator $\fone$ and $\kone$
of the pseudoscalar and vector channel, respectively.
The definitions are such that, at the $\minv$ order of the heavy quark expansion,
the observables assume the following form 
\bea
\Phi_1^{\rm HQET} &=& L \big(\mhbare +  \meffstat +  \cahqet\meffdastat + \omegakin\meffkin + \omegaspin\meffspin \big) \,,\\
\Phi_2^{\rm HQET} &=& \lnzahqet + \zetaa +  \cahqet\rhod + \omegakin\psikin + \omegaspin\psispin \,,\\
\Phi_3^{\rm HQET} &=& \rastat +  \cahqet\rda + \omegakin\rakin + \omegaspin\raspin \,,\\
\Phi_4^{\rm HQET} &=& \ronestat + \omegakin\ronekin \,,\\
\Phi_5^{\rm HQET} &=& \omegaspin\rhoonespin \,,
\eea
which is just an explicit version of \eq{eq:phihqet}.
The precise definitions of the HQET quantities can be found in~\cite{Blossier:2010jk}.

\section{Tuning of $L_1$ and renormalization in finite-volume QCD}

\label{a:tuning}
As explained in the main text, the basic element of our non-perturbative
strategy to compute the HQET parameters consists in imposing matching
conditions between a set of renormalized finite-volume observables in QCD, 
extrapolated to the continuum limit, and their counterparts in HQET, 
expanded up to order $1/\mh$.
Since in this way the HQET parameters get determined by feeding results from
non-perturbatively renormalized QCD into the effective theory, we  
summarize here how the renormalization in QCD (i.e., of the gauge coupling, 
the quark masses and the relevant composite fields) is performed.

We work in a small volume of
linear extent $L_1\approx 0.4\,\Fm$, where the SF 
setup~\cite{Luscher:1992an,Sint:1993un}, with $T=L$ and 
$\theta=0.5$ as the periodicity angle of the sea quark fields, serves as our 
finite-volume renormalization scheme.
The physical volume stands in one-to-one correspondence to 
the SF gauge coupling $\gbsq(L)$ running with the 
scale $L$~\cite{Luscher:1993gh,DellaMorte:2004bc}. 
We thus define $L_1$ by the condition
\be
\gbsq(L_1/2)=2.989 \;.
\label{eqn:cond-L1}
\ee
The known step scaling function of the SF coupling in two-flavour 
QCD~\cite{DellaMorte:2004bc} implies
\be
\gbsq(L_1)=\sigma(2.989)=4.484(48) \;,
\label{eqn:ssf-L1}
\ee
and the associated exact value of $L_1$ in physical units could then be 
inferred from the results of~\cite{Marinkovic:2011pa}, but it is not relevant 
for the following.

Our finite-volume QCD observables $\Phi_i^{\rm QCD}$, $i=1,\ldots,5$, are defined as 
suitable renormalized combinations of SF correlation functions (see \ref{a:observables})
composed of a non-degenerate heavy-light valence quark doublet, where the light valence 
quark mass is chosen to be equal to the mass $\ml$ of the mass degenerate 
dynamical sea quark doublet.
The hopping parameter of the corresponding light valence ($=$ sea) 
quark is denoted as $\kapl$ and that of the heavy valence quark by $\kaph$.
The $\Phi_i^{\rm QCD}$ are universal and, in particular, their continuum limits exist,
once they have been evaluated in numerical simulations along a line of
constant physics specified by a series of bare parameters 
$(L/a,\beta,\kapl,\kaph)$ such that the renormalized SF coupling and the 
light and heavy quark masses are kept fixed. 

According to eq.~(\ref{eqn:cond-L1}), $\beta$ is determined by requiring
$\gbsq(L_1/2)=2.989$ for given resolutions $2a/L_1$.
This peculiar value of the SF coupling was indicated by an initial 
simulation with $L_1/(2a)=20$, while for $10\le L_1/(2a)\le 16$ additional 
simulations and interpolations in $\beta$, based on the known dependence of 
the SF coupling and the sea quark mass on the bare parameters 
available from the data of~\cite{DellaMorte:2004bc}, were employed for 
fine-tuning to that target.
Employing the non-perturbative $\beta$-function of the SF coupling and 
estimates from our quenched calculation on the effect of propagating 
uncertainties in $\gbsq$ (cf.~Appendix~D of~\cite{DellaMorte:2006cb}), 
we can assess an uncertainty of about $0.05$ in $\gbsq$ to translate via 
$L_1$ into an uncertainty in the b-quark mass of at most $0.5\%$, 
which is negligible compared to the present direct uncertainty in the quark mass 
renormalization discussed below.

In the quark sector, the sea and light valence quark masses are taken to 
be the same; in the numerical computation they are actually 
tuned to zero.
The condition $\ml(L_1/2)=0$ is met by setting $\kapl$ to the critical 
hopping parameter, $\kapc$, which is determined by the vanishing of the PCAC 
mass of the sea quark doublet, defined as in eq.~(2.17) 
of~\cite{DellaMorte:2004bc} through the $\Or(a)$-improved axial current in the 
SF setup with boundary field~"A"~\cite{Luscher:1993gh}.
Again, $\kapc$ was estimated and partly fine-tuned on basis of the data 
published in~\cite{DellaMorte:2004bc}, whereby for the improvement 
coefficient of the axial current, $\ca$, 1-loop perturbation 
theory~\cite{Luscher:1996vw} and the non-perturbative estimates 
of~\cite{DellaMorte:2005se} (after they had become available) were used.  
A slight mismatch of $|\frac{L_1}{2}\,\ml(L_1/2)|<0.05$ of this condition is 
tolerable in practice.  
The resulting triples $(L_1/a,\beta,\kapl)$ are collected in the three
leftmost columns of Table~\ref{tab:L1QCDparam}. 
Note that the mentioned 1-loop value for $\ca$ was only used in the 
preliminary determination of $\kapl$. All PCAC masses listed here 
are computed with the non-perturbative $\ca$ of~\cite{DellaMorte:2005se}.
%
\TABLES{%
\small
\renewcommand{\arraystretch}{1.25}
\begin{tabular}{ccccccccc} 
\toprule
  $L/a$ & $\beta$ & $\kapl$ & $\gbsq\big(\frac{L}{2}\big)$ 
& $\zp\big(g_0,\frac{L}{2a}\big)$ & $\bm$ & $Z$ & $z$ & $\kaph$ \\\midrule
  $20$ & $6.1569$ & $0.1360536$ & $2.989(36)$ & $0.6065(9)$  & $-0.6633(12)$ & $1.10443(17)$ & $4$  & $0.1327094$  \\[-0.4em]
        &&&&&&& $6$  & $0.1309180$  \\[-0.4em]
        &&&&&&& $7$  & $0.1299824$  \\[-0.4em]
        &&&&&&& $9$  & $0.1280093$  \\[-0.4em]
        &&&&&&& $11$ & $0.1258524$  \\[-0.4em]
        &&&&&&& $13$ & $0.1234098$  \\[-0.4em]
        &&&&&&& $15$ & $0.1204339$  \\[-0.4em]
        &&&&&&& $18$ & $\text{---}$ \\[-0.4em]
        &&&&&&& $21$ & $\text{---}$ \\[0.2em]
  $24$ & $6.2483$ & $0.1359104$ & $2.989(30)$ & $0.5995(8)$  & $-0.6661(9)$  & $1.10475(12)$ & $4$  & $0.1331966$  \\[-0.4em]
        &&&&&&& $6$  & $0.1317649$  \\[-0.4em] 
        &&&&&&& $7$  & $0.1310257$  \\[-0.4em]
        &&&&&&& $9$  & $0.1294907$  \\[-0.4em]
        &&&&&&& $11$ & $0.1278628$  \\[-0.4em]
        &&&&&&& $13$ & $0.1261106$  \\[-0.4em]
        &&&&&&& $15$ & $0.1241815$  \\[-0.4em]
        &&&&&&& $18$ & $0.1206988$  \\[-0.4em]
        &&&&&&& $21$ & $0.1140810$  \\[0.2em]
  $32$ & $6.4574$ & $0.1355210$ & $2.989(35)$ & $0.5941(10)$ & $-0.6674(23)$ & $1.10455(17)$ & $4$  & $0.1335537$  \\[-0.4em]
        &&&&&&& $6$  & $0.1325329$  \\[-0.4em] 
        &&&&&&& $7$  & $0.1320117$  \\[-0.4em]
        &&&&&&& $9$  & $0.1309446$  \\[-0.4em]
        &&&&&&& $11$ & $0.1298401$  \\[-0.4em]
        &&&&&&& $13$ & $0.1286909$  \\[-0.4em]
        &&&&&&& $15$ & $0.1274876$  \\[-0.4em]
        &&&&&&& $18$ & $0.1255509$  \\[-0.4em]
        &&&&&&& $21$ & $0.1233865$  \\[0.2em]
  $40$ & $6.6380$ & $0.1351923$ & $2.989(43)$ & $0.5949(12)$ & $-0.6692(27)$ & $1.10379(17)$ & $4$  & $0.1336432$  \\[-0.4em]
        &&&&&&& $6$  & $0.1328462$  \\[-0.4em]
        &&&&&&& $7$  & $0.1324413$  \\[-0.4em]
        &&&&&&& $9$  & $0.1316178$  \\[-0.4em]
        &&&&&&& $11$ & $0.1307738$  \\[-0.4em]
        &&&&&&& $13$ & $0.1299065$  \\[-0.4em]
        &&&&&&& $15$ & $0.1290126$  \\[-0.4em]
        &&&&&&& $18$ & $0.1276125$  \\[-0.4em]
        &&&&&&& $21$ & $0.1261232$  \\
\bottomrule   
\end{tabular}
\caption{\footnotesize
Bare parameters $(L/a,\beta,\kapl,\kaph)$ used in the computation of the 
heavy-light QCD observables for $L=L_1$.
As explained in the text, they are fixed by the renormalization conditions
$\gbsq(L/2)=2.989$, $\frac{L}{2}\,\ml(L/2)\approx 0$ and $LM=z$ for the SF 
coupling and the light PCAC and heavy RGI quark masses, respectively.
The entering renormalization constants $\zp$ and $Z$ and the improvement 
coefficient $\bm$ were calculated in~\protect\cite{Fritzsch:2010aw}.
}\label{tab:L1QCDparam}

}
%

It remains to fix the renormalized mass of the heavy valence quark to a
sequence of values, fairly spanning a range from around the charm to beyond 
the bottom quark mass.
To this end we choose the dimensionless variable $z\equiv LM$, with $M$ 
being the RGI mass of the heavy valence 
quark, because the latter is related via
\be
M=
h(L)\,\zm(g_0,L/a)\,\left(1+\bm(g_0)\,a\mqh\right)\,\mqh
\,+\,\Or(a^2)
\label{eqn:M-mqtil}
\ee
to the subtracted bare heavy quark mass, $\mqh$, and its hopping parameter, 
$\kaph$, in the $\Or(a)$-improved theory.
Here,
\be
\zm(g_0,L/a)=
\frac{Z(g_0)\,\za(g_0)}{\zp(g_0,L/a)} \;,\quad
a\mqh=
\frac{1}{2}\left(\frac{1}{\kaph}-\frac{1}{\kapc}\right) \;,\quad
L=L_1/2 \;.
\label{eqn:Zm-mqh}
\ee
All ingredients of eqs.~(\ref{eqn:M-mqtil}) and (\ref{eqn:Zm-mqh}) are
non-perturbatively known in two-flavor QCD:
the axial current renormalization constant $\za$ from ~\cite{DellaMorte:2008xb}, 
and the renormalization factor $Z$ and the improvement coefficient $\bm$ 
from~\cite{Fritzsch:2010aw}. 
The scale dependent renormalization constant $\zp$ for the specific
$\beta$-values in question was extracted in~\cite{Fritzsch:2010aw}, following 
exactly the definition of~\cite{DellaMorte:2005kg}.
$\zp$, $\bm$ and $Z$ are also listed in Table~\ref{tab:L1QCDparam}.
In eq.~\eqref{eqn:M-mqtil}, there also appears the factor
\be
h(L)\equiv
\frac{M}{\mbar(\mu)}=1.521(14) \;,\quad
\mu=
1/L=2/L_1 \;,
\label{eqn:h-L0}
\ee
which represents the universal, regularization independent ratio of the RGI 
heavy quark mass to the running quark mass, $\mbar$, in the SF scheme at 
the renormalization scale $\mu$. 
$h(L_1/2)$ was evaluated by a reanalysis of available data on the 
non-perturbative quark mass renormalization in two-flavour QCD as published 
in~\cite{DellaMorte:2005kg}.

Given the values 
\be
L_1M=z\in\{4,6,7,9,11,13,15,18,21\}
\label{eqn:fixed-z-values}
\ee
of the dimensionless RGI heavy quark mass in $L_1$ and resolutions
$L_1/a=20,24,32,40$, eqs.~\eqref{eqn:M-mqtil} -- \eqref{eqn:h-L0} can 
now straightforwardly be solved for the corresponding nine heavy valence 
quark hopping parameters $\kaph=\kaph(z,g_0)$ that fix $z$ to the numbers 
in eq.~\eqref{eqn:fixed-z-values}.%
\footnote{%
Owing to the sign and the order of magnitude of the non-perturbative values 
for $\bm$ in the $\beta$-range relevant here, $\kaph(z,g_0)$ has no real 
solutions for arbitrarily high $z$-values.
This implies that only for inverse lattice spacings $L_1/a=24,32,40$
hopping parameters $\kaph$ that achieve $z=18,21$ can be found.
}
These hopping parameters and the associated $z$-values are collected in the
two rightmost columns of Table~\ref{tab:L1QCDparam}.

Table~\ref{tab:L1QCDparam} lists the bare parameters $L/a$, $\beta$, 
$\kapl$, $\kaph^{(i)}$ ($i=1,\ldots,9$) of the numerical 
simulations in $L=L_1$, from which the heavy-light QCD observables $\Phi_i$, 
$i=1,\ldots,5$, are computed.
These bare parameters are functions of the dimensionless variables $\gbsq(L)$, 
$z=LM$ and the resolution $a/L$, with well-defined continuum limits at fixed
$z$, but we can also consider them as functions of the box size $L$, the RGI 
mass $M$ of the heavy quark and the lattice spacing $a$.

Let us still comment on the error budget arising from the above procedure of 
fixing~$z$~\cite{Fritzsch:2010aw}, which has to be accounted for in any
secondary quantity analyzed as a function of~$z$.
From the uncertainties on $\za$, $\zp$, $Z$, and $\bm$ quoted in the
respective references~\cite{DellaMorte:2008xb,Fritzsch:2010aw} (see also
Table~\ref{tab:L1QCDparam}), one obtains by the standard rules of Gaussian 
error propagation an accumulated relative error on $z$ in the range 
$0.38\%\le(\Delta z/z)\le 0.41\%$ for all $z$-values and lattice resolutions
in use here.
The contribution from the universal continuum factor $h(L_1/2)$,
\eq{eqn:h-L0}, represents with $\Delta h/h=0.92\%$ the dominating 
source of uncertainty in the total error budget of $\Delta z/z=1.01\%$.
Note, however, that the error on this universal factor $h$ has to be 
propagated into the QCD observables $\Phi_i$ only after their extrapolations
to the continuum limit. 
It is not included in the errors in Table~\ref{tab:L1QCDparam}.

A similar tuning procedure has to be performed for the HQET simulations 
necessary for the matching. Since the matching proceeds through renormalized
quantities and therefore in the continuum limit, there is no need to 
use the same lattice resolutions on both the HQET and QCD sides.
Larger lattice spacings can obviously be chosen in HQET compared to the 
relativistic case, as long as the choices are such that the condition in 
eq.~(\ref{eqn:cond-L1}) is fulfilled.
The resulting bare parameters and values of $\gbsq(L)$ and the bare PCAC sea
quark mass, defined as before, are collected in Table~\ref{tab:L1HQETparam}.
%
\TABLES{%
\begin{center}
\small
\renewcommand{\arraystretch}{1.25}
\begin{tabular}{ccccc}
\toprule
    $L/a$ & $\beta $ & $\kapl$ & $\gbsq(L)$ & $a\ml$ \\ 
\midrule
    $ 6 $        & $5.2638$ & $0.135985 $ & $4.423(75)$             &  $-0.01154(83)$   \\
    $ 8 $        & $5.4689$ & $0.136700 $ & $4.473(83)$             &  $-0.00424(24)$   \\
    $ 10$        & $5.6190$ & $0.136785 $ & $4.49(10) $\hphantom{0} &  $-0.00257(11)$   \\
    $ 12$        & $5.7580$ & $0.136623 $ & $4.501(91)$             &  $+0.00067(7)$\hphantom{0}    \\
    $ 16$        & $5.9631$ & $0.136422 $ & $4.40(10) $\hphantom{0} &  $-0.00096(4)$\hphantom{0}    \\\cmidrule(lr){1-5} 
    $8^{\ast}$   & $5.4689$ & $0.13564  $ & $4.873(99)$             &  $+0.03189(18)$  \\
    $12^{\ast}$  & $5.8120$ & $0.136617 $ & $4.218(49)$             &  $-0.00099(7)$\hphantom{0}    \\
\bottomrule
\end{tabular}
\caption{
\footnotesize
Bare parameters and results of the tuning to $\gbsq(L)=4.484$ for the HQET simulations
entering the matching step. The additional lattices $L/a=8^{\ast},12^{\ast}$ are used
to estimate and propagate a potential error, resulting from not meeting the line of
constant physics condition exactly.}
 \label{tab:L1HQETparam}
\end{center}

}
%
From the two runs at $L/a=8$ we estimate
$s=\left.{{\partial \gbar^2}\over{\partial z}}\right|{\!}_{\gbar^2=4.484}=1.4(4)$,
which we use to set a bound on the quark mass $\ml$ (which should be vanishing), 
such that its effect on the coupling is below the statistical error.
The $L/a=12^{\ast}$ label refers to an additional run used to propagate 
the error on $\gbar^2$ into the HQET observables. This is done by computing
all HQET observables (at $T=L$ and $T=L/2$) also at the bare parameters of $12^\ast$ 
in order to estimate the variation of our primary HQET observables with 
respect to a variation of the renormalized coupling. This procedure 
neglects the lattice spacing dependence of the variation,
which is justified for an error computation.
In practice, the uncertainty on $\gbsq(L)$ is the dominating
piece of the errors of $\eta_3,\,\eta_4$ shown in \fig{fig:PhiRA1stat},
while it can safely be neglected for all other observables 
within our present error budget.

\section{Simulation details}
\label{a:sim_details}
Our computations are in a natural way split
into two parts, the generation of $\Nf=2$ gauge field ensembles at the tuned
parameters $(L/a,\beta,\kappa_{\rm sea}\equiv\kapl)$, and the subsequent
computation of all relevant correlation functions. The Sheikoleslami-Wohlert
improvement coefficient $\csw(g_0)$ is set to its non-perturbative
estimate \cite{Jansen:1998mx} for $\nf=2$.  In order to have an
improved action in the SF we
include boundary counterterms to cancel boundary-induced lattice artefacts.
The corresponding improvement coefficients, $\ct(g_0)$ and $\cttil(g_0)$, 
are always set to their known 2-loop~\cite{Bode:1999sm} and
1-loop~\cite{Luscher:1996vw} values, respectively.  This guarantees
that any boundary-to-boundary correlation function, such as 
$f_1$, is $\Or(a)$-improved.
\vskip1em
\noindent{\bf Ensemble generation:} For simulating a doublet of mass degenerate,
non-perturbatively improved dynamical Wilson fermions in the SF $(\theta_{\rm sea}=0.5)$,
we use the algorithmic implementation described in 
detail in~\cite{DellaMorte:2008ad} to produce 
the ensembles needed for the matching in the volume $L_1$.
Applying the step scaling technique in HQET to volume $L_2=2L_1$ also requires simulations
at resolutions $a/(2L_1)$ while keeping all other parameters fixed.
Furthermore, our strategy involves simulations at fixed time extent
$T=L$ as well as $T=L/2$.

Since most of the simulations used for the HQET observables are fast and can 
usually be
performed using several replica, we aimed for a total statistic of at least
8000 configurations in these ensembles. Only for our most expensive HQET
ensembles with $L_2/a=32$ we did not reach this goal due to limited computing
resources and thus restricted ourselves to have $\Or(3000)$ configurations
here. 
The QCD simulations have larger values of $L/a$ (and thus smaller lattice 
spacings) compared to the HQET simulations. It is therefore more difficult 
to achieve high statistics for the QCD ensembles.
Even more
so since the gap in the spectrum of the Dirac operator, 
which allows to simulate at vanishing quark
mass in the SF, decreases proportionally to the inverse time extent. Hence, for
the production of ensembles used to measure QCD observables, our goal 
was just
to reach a reasonable statistics, and thereby to obtain small and comparable errors in our
final observables at the different resolutions. Thus, the ensemble size
roughly increases from $\Or(500)$ at $L/a=20$ to $\Or(1500)$ at $L/a=40$.

Using the notation introduced in~\cite{DellaMorte:2008ad}, we list the relevant
algorithmic parameters and additional details in \Tab{tab:algopar}.  For QCD
the bare parameters $(L/a,\beta,\kapl)$ are those of Table~\ref{tab:L1QCDparam},
while for HQET we use those in Table~\ref{tab:L1HQETparam} which are not marked
by a star.  The molecular dynamics (MD) is characterized by specifying
the trajectory length, the integrator, and the step size(s).
For the trajectory length we choose $\tau=2$ in MD units since we expect 
autocorrelation to be reduced~\cite{Meyer:2006ty} in that case. 
As integration scheme we always use multiple time scales with leap-frog
integrator, also known as Sexton-Weingarten scheme~\cite{Sexton:1992nu}.
The tunable algorithmic parameter $\rho_0$ is introduced
as a mass-preconditioning of the Dirac-operator \`a la Hasenbusch~\cite{algo:GHMC}.
The step sizes for the corresponding two pseudofermions are $\delta\tau_0$
and $\delta\tau_1$, while for
the gauge force we use
$\delta\tau_0/\delta\tau_g = 4$ throughout. $\langle N_{\rm
CG}^{(i)}\rangle$ is the average number of conjugate-gradient iterations used
to solve the symmetrically even-odd preconditioned Dirac equation during the
trajectory, and  
$P_{\rm acc}$ is the acceptance rate of the
simulation. $\tau_{\rm meas}$ gives the MD time between configurations which have
been stored on disk and used for measurements.
In case we list more than one value of $\tau_{\rm meas}$,
we have performed independent simulations with different measurement frequencies which
have been chosen to be of the typical size of observed integrated
autocorrelation times $\tau_{\rm int}$.  Explicitly, we show the average
plaquette and PCAC mass
of the production runs together with
their estimated autocorrelation time in Table~\ref{tab:algores}. 
We also list the results for the pseudoscalar boundary-to-boundary correlation
function $f_1$, which typically is the quantity with the largest integrated
autocorrelation time among the different SF correlation functions.
\TABLES{%
\renewcommand{\arraystretch}{1.25}
\begin{tabular}{cccccccccc} 
\toprule
sector       & $L/a$ & $T/a$ & [$\tfrac{\tau}{\delta\tau_0},\tfrac{\delta\tau_0}{\delta\tau_1}$] %
                                               & $\rho_0$   & $\langle N_{\rm CG}^{(0)}\rangle$ %
                                                                    & $\langle N_{\rm CG}^{(1)}\rangle$ %
                                                                              & $P_{\rm acc}$ & $\langle {\rm e}^{-\Delta H}\rangle$ %
                                                                                          & $\tau_{\rm meas}$  \\\midrule       %
QCD[$L_1$]   & $20$ & $20$  & [\,$40,\,4$\,]   &  $0.0828$ & $51$   &  $292$  &  $87\%$   & $0.9896(84)$ &  $8$       \\[-0.2em]               
$\downarrow$ & $24$ & $24$  & [\,$50,\,5$\,]   &  $0.0755$ & $56$   &  $346$  &  $90\%$   & $1.0006(45)$ &  $10$      \\[-0.2em]               
             & $32$ & $32$  & [\,$56,\,4$\,]   &  $0.0651$ & $63$   &  $442$  &  $88\%$   & $1.0009(48)$ &  $10$      \\[-0.2em]               
             & $40$ & $40$  & [\,$64,\,5$\,]   &  $0.0450$ & $81$   &  $533$  &  $90\%$   & $0.9986(42)$ &  $4$       \\\cmidrule(lr){2-10}    
             & $20$ & $10$  & [\,$40,\,4$\,]   &  $0.0828$ & $50$   &  $145$  &  $93\%$   & $1.0012(29)$ &  $8 $      \\[-0.2em]               
             & $24$ & $12$  & [\,$46,\,5$\,]   &  $0.1140$ & $39$   &  $166$  &  $89\%$   & $0.9991(56)$ &  $10$      \\[-0.2em]               
             & $32$ & $16$  & [\,$48,\,4$\,]   &  $0.0977$ & $45$   &  $211$  &  $84\%$   & $0.9990(83)$ &  $8 $      \\[-0.2em]               
             & $40$ & $20$  & [\,$54,\,4$\,]   &  $0.0870$ & $49$   &  $257$  &  $84\%$   & $0.9983(80)$ &  $6 $      \\\midrule               
HQET[$L_1$]  & $ 6$ &  $ 6$ & [\,$30,\,4$\,]   &  $0.150$  & $27$   &  $106$  &  $92\%$   & $0.9989(8) $ &  $10$      \\[-0.2em]               
$\downarrow$ & $ 8$ &  $ 8$ & [\,$30,\,4$\,]   &  $0.130$  & $35$   &  $137$  &  $90\%$   & $0.9983(13)$ &  $10$      \\[-0.2em]               
             & $10$ &  $10$ & [\,$30,\,5$\,]   &  $0.110$  & $32$   &  $137$  &  $89\%$   & $1.0006(14)$ &  $10$      \\[-0.2em]               
             & $12$ &  $12$ & [\,$32,\,4$\,]   &  $0.100$  & $44$   &  $191$  &  $89\%$   & $0.9990(14)$ &  $10$      \\[-0.2em]               
             & $16$ &  $16$ & [\,$40,\,4$\,]   &  $0.085$  & $50$   &  $247$  &  $91\%$   & $1.0021(12)$ &  $10$      \\\cmidrule(lr){2-10}    
             & $ 6$ &  $ 3$ & [\,$24,\,4$\,]   &  $0.243$  & $21$   &  $47 $  &  $91\%$   & $1.0004(12)$ &  $8 $      \\[-0.2em]               
             & $ 8$ &  $ 4$ & [\,$28,\,4$\,]   &  $0.206$  & $24$   &  $63 $  &  $90\%$   & $1.0002(12)$ &  $10$      \\[-0.2em]               
             & $10$ &  $ 5$ & [\,$30,\,5$\,]   &  $0.182$  & $26$   &  $77 $  &  $89\%$   & $1.0016(15)$ &  $8 $      \\[-0.2em]               
             & $12$ &  $ 6$ & [\,$32,\,4$\,]   &  $0.166$  & $28$   &  $90 $  &  $89\%$   & $1.0003(15)$ &  $10$      \\[-0.2em]               
             & $16$ &  $ 8$ & [\,$40,\,4$\,]   &  $0.141$  & $33$   &  $118$  &  $90\%$   & $1.0003(13)$ &  $10$      \\\midrule               
HQET[$L_2$]  & $12$ &  $12$ & [\,$52,\,5$\,]   &  $0.100$  & $46$   &  $256$  &  $92\%$   & $1.0018(32)$ &  $6;10$    \\[-0.2em]               
$\downarrow$ & $16$ &  $16$ & [\,$50,\,4$\,]   &  $0.085$  & $52$   &  $327$  &  $90\%$   & $0.9999(15)$ &  $6;10$    \\[-0.2em]               
             & $20$ &  $20$ & [\,$54,\,4$\,]   &  $0.081$  & $54$   &  $394$  &  $88\%$   & $1.0017(27)$ &  $6$       \\[-0.2em]               
             & $24$ &  $24$ & [\,$50,\,4$\,]   &  $0.070$  & $61$   &  $441$  &  $86\%$   & $1.0003(23)$ &  $6;10$    \\[-0.2em]               
             & $32$ &  $32$ & [\,$64,\,4$\,]   &  $0.063$  & $71$   &  $747$  &  $88\%$   & $1.0059(78)$ &  $8;10$    \\\cmidrule(lr){2-10}    
             & $12$ &  $6$  & [\,$40,\,4$\,]   &  $0.166$  & $30$   &  $114$  &  $89\%$   & $1.0010(14)$ &  $6;10$    \\[-0.2em]               
             & $16$ &  $8$  & [\,$40,\,4$\,]   &  $0.141$  & $34$   &  $144$  &  $92\%$   & $0.9996(25)$ &  $6;10$    \\[-0.2em]               
             & $20$ &  $10$ & [\,$50,\,4$\,]   &  $0.100$  & $44$   &  $177$  &  $87\%$   & $1.0006(13)$ &  $6$       \\[-0.2em]               
             & $24$ &  $12$ & [\,$50,\,4$\,]   &  $0.114$  & $40$   &  $198$  &  $85\%$   & $0.9986(20)$ &  $6;10$    \\[-0.2em]               
             & $32$ &  $16$ & [\,$52,\,4$\,]   &  $0.091$  & $48$   &  $256$  &  $86\%$   & $0.9921(45)$ &  $8;10$    \\                       
\bottomrule   
\end{tabular}
\caption{%
Algorithmic parameters of our production runs as explained in the text.
}\label{tab:algopar}

}
\vskip1em
\noindent{\bf Measurements:} 
since there is no explicit heavy quark mass contribution to correlation functions in
HQET, we just need to specify $\kappa_{\rm l}\equiv\kappa_{\rm sea}$ and the
static quark action(s) in use to compute static-light correlation functions.
The latter have been computed using the two static actions HYP1,2 described
in~\cite{Della Morte:2005yc}.
For measurements in QCD we compute heavy-light observables in a partially
quenched setup with $\kappa_{\rm l}\equiv\kappa_{\rm sea}$ and heavy valence
quark hopping parameters $\kappa_{\rm val,h}\equiv\kappa_{\rm h}^{(i)}$,
$i=1,\ldots,9$. 
The latter slightly deviate from those in Table~\ref{tab:L1HQETparam} because
at the time we fixed the values of $z$ the updated $g_0^2$-dependence of 
$\za$~\cite{DellaMorte:2008xb} entering eq.~\eqref{eqn:M-mqtil} was not yet at hand. 
This translates into a small mismatch $(\lesssim 0.4\%)$ of the $z$-values where we
did the computation with respect to our target $z$-values. The QCD observables
were interpolated to the target $z$-values to account for this.
%
\TABLES{%
\renewcommand{\arraystretch}{1.25}
\begin{tabular}{ccccccccc} 
\toprule
sector       & $L/a$ & $T/a$ & $\langle$plaquette$\rangle$ & $\tau_{\rm int}[\text{plaq}]$ & $\langle am\rangle$ & $\tau_{\rm int}[am]$ & $\langle f_1\rangle$ & $\tau_{\rm int}[f_1]$  \\\midrule
QCD[$L_1$]   & $20$  & $20$  & $0.630066(13)$ & $3.6(6)$ & $+0.00055(13) $ & $8(2)  $  & $0.4582(69)$  & $24(8)  $  \\[-0.2em]
$\downarrow$ & $24$  & $24$  & $0.636756(8)$  & $4.7(7)$ & $-0.000145(66)$ & $7(1)  $  & $0.434(11) $  & $80(30) $  \\[-0.2em]
             & $32$  & $32$  & $0.651034(3)$  & $4.1(5)$ & $+0.000146(32)$ & $5.4(7)$  & $0.4155(72)$  & $60(20) $  \\[-0.2em]
             & $40$  & $40$  & $0.662409(2)$  & $3.9(5)$ & $+0.000034(17)$ & $2.7(3)$  & $0.3990(98)$  & $100(40)$  \\\cmidrule(lr){2-9}
             & $20$  & $10$  & $0.629675(14)$ & $4.8(6)$ & $+0.000182(57)$ & $4.4(5)$  & $0.9474(16)$  & $8(1)   $  \\[-0.2em]
             & $24$  & $12$  & $0.636422(11)$ & $5.0(7)$ & $-0.000400(56)$ & $5.4(7)$  & $0.9280(19)$  & $9(2)   $  \\[-0.2em]
             & $32$  & $16$  & $0.650779(7)$  & $5.0(9)$ & $+0.000001(36)$ & $4.1(5)$  & $0.8892(43)$  & $40(10) $  \\[-0.2em]
             & $40$  & $20$  & $0.662203(4)$  & $5.6(9)$ & $-0.000056(22)$ & $3.8(5)$  & $0.8769(29)$  & $21(6)  $  \\\midrule
HQET[$L_1$]  & $ 6$  &  $ 6$ & $0.546135(43)$ & $5.5(2)$ & $-0.00585(18)$  & $5.9(2)$  & $0.5041(11)$  & $5.9(2) $  \\[-0.2em]
$\downarrow$ & $ 8$  &  $ 8$ & $0.569268(25)$ & $5.1(3)$ & $-0.00339(13)$  & $5.9(3)$  & $0.5128(12)$  & $6.2(3) $  \\[-0.2em]
             & $10$  &  $10$ & $0.584467(14)$ & $4.6(2)$ & $-0.00260(9)$   & $5.7(3)$  & $0.5010(13)$  & $8.7(5) $  \\[-0.2em]
             & $12$  &  $12$ & $0.597377(10)$ & $5.1(2)$ & $+0.00040(6)$   & $5.4(2)$  & $0.4599(13)$  & $12.3(9)$  \\[-0.2em]
             & $16$  &  $16$ & $0.615017(5)$  & $4.9(2)$ & $-0.00107(4)$   & $5.7(3)$  & $0.4676(15)$  & $18(2)  $  \\\cmidrule(lr){2-9}
             & $ 6$  &  $ 3$ & $0.547894(68)$ & $4.6(2)$ & $-0.01686(11)$  & $4.7(2)$  & $1.1141(11)$  & $5.1(3) $ \\[-0.2em]
             & $ 8$  &  $ 4$ & $0.569319(34)$ & $4.6(2)$ & $-0.02096(10)$  & $5.4(2)$  & $1.0918(9) $  & $5.6(2) $ \\[-0.2em]
             & $10$  &  $ 5$ & $0.584064(23)$ & $4.7(3)$ & $-0.01167(7)$   & $4.5(2)$  & $1.0585(8) $  & $4.7(2) $ \\[-0.2em]
             & $12$  &  $ 6$ & $0.596924(14)$ & $5.0(3)$ & $-0.00338(5)$   & $5.6(3)$  & $1.0037(7) $  & $6.1(3) $ \\[-0.2em]
             & $16$  &  $ 8$ & $0.614569(7)$  & $4.5(2)$ & $-0.00226(3)$   & $5.7(3)$  & $0.9911(6) $  & $6.5(4) $ \\\midrule
HQET[$L_2$]  & $12$  &  $12$ & $0.546446(16)$ & $6.8(4)$ & $+0.00798(14)$  & $7.7(5)$  & $0.2527(17)$  & $19(2)  $ \\[-0.2em]
$\downarrow$ & $16$  &  $16$ & $0.569646(8)$  & $5.2(3)$ & $+0.000457(65)$ & $4.8(3)$  & $0.3044(26)$  & $33(4)  $ \\[-0.2em]
             & $20$  &  $20$ & $0.584826(6)$  & $4.9(4)$ & $-0.000900(55)$ & $4.3(3)$  & $0.3057(44)$  & $41(8)  $ \\[-0.2em]
             & $24$  &  $24$ & $0.597708(3)$  & $4.7(3)$ & $+0.001342(32)$ & $6.2(4)$  & $0.2650(28)$  & $65(15) $ \\[-0.2em]
             & $32$  &  $32$ & $0.615288(5)$  & $3.6(6)$ & $-0.000841(61)$ & $4.6(8)$  & $0.280(18) $  & $130(60)$ \\\cmidrule(lr){2-9}
             & $12$  &  $6$  & $0.547364(19)$ & $5.9(3)$ & $-0.003649(87)$ & $6.7(4)$  & $0.8076(9) $  & $7.4(4) $ \\[-0.2em]
             & $16$  &  $8$  & $0.569606(12)$ & $5.4(4)$ & $-0.002321(63)$ & $5.5(3)$  & $0.8047(10)$  & $6.6(5) $ \\[-0.2em]
             & $20$  &  $10$ & $0.584593(6)$  & $4.6(3)$ & $-0.001945(33)$ & $4.1(2)$  & $0.7840(10)$  & $10.0(9)$ \\[-0.2em]
             & $24$  &  $12$ & $0.597446(4)$  & $4.9(3)$ & $+0.000785(23)$ & $4.9(2)$  & $0.7120(9) $  & $14(1)  $ \\[-0.2em]
             & $32$  &  $16$ & $0.615045(5)$  & $5.3(6)$ & $-0.000923(29)$ & $5.0(5)$  & $0.7322(25)$  & $29(6)  $ \\
\bottomrule   
\end{tabular}
\caption{%
Results for standard quantities measured during the production runs.
}\label{tab:algores}

}

\end{appendix}

\clearpage
\bibliographystyle{JHEP}   
\bibliography{param}        

\end{document}